\documentclass[12pt,a4paper]{article}
\pdfoutput=1
\usepackage[a4paper]{geometry}
\usepackage{graphicx}
\usepackage{amsmath}
\usepackage{amssymb}
\usepackage{float}
\usepackage{listings}
\newcommand{\asb}{\bar{\alpha}_s}

\newcommand{\stringa}{\ttfamily\lstinline}
\def\cod#1{{\stringa!#1!}}

\title{\bf BFKL Azimuthal Imprints in Inclusive Three-jet Production at 7 and 13 TeV}
\author{F. Caporale$^1$, F.~G. Celiberto$^{1,2}$, G. Chachamis$^1$, \\ 
        D. Gordo G{\' o}mez$^1$, A. Sabio Vera$^{1}$\\ \\
{\small $^1$ Instituto de F{\' \i}sica Te{\' o}rica UAM/CSIC, Nicol{\'a}s Cabrera 15}\\ 
{\small \& Universidad Aut{\' o}noma de Madrid, E-28049 Madrid, Spain.}\\
{\small $^2$ Dipartimento di Fisica, Universit{\`a} della Calabria \&}\\
{\small Istituto Nazionale di Fisica Nucleare, Gruppo Collegato di Cosenza,}\\
{\small I-87036 Arcavacata di Rende, Cosenza, Italy.}
}

\begin{document}

\maketitle 

\abstract
We propose the study of new observables in LHC inclusive events with three tagged jets, one in the forward direction, one in the backward direction and both well-separated in rapidity
from the each other (Mueller-Navelet jets), together with a third jet tagged in central regions of rapidity. Since non-tagged associated mini-jet multiplicity is allowed, we argue that  projecting the cross sections on azimuthal-angle components can provide several distinct tests of the BFKL dynamics. Realistic LHC kinematical cuts are introduced.

\section{Introduction}

In recent years the steady running of the Large Hadron Collider (LHC) has opened up new avenues for the study of the high energy limit of Quantum Chromodynamics (QCD). This is particularly important in the context of jet production since the abundance of data allows for the possibility to study more exclusive observables, needed to isolate regions of phase hidden in more inclusive, previous, analysis. In this work we focus on the investigation of jet production in the so-called multi-Regge kinematics. When jets are produced at large relative rapidities the Balitsky-Fadin-Kuraev-Lipatov (BFKL) approach in the leading logarithmic (LL)~\cite{Lipatov:1985uk,Balitsky:1978ic,Kuraev:1977fs,Kuraev:1976ge,Lipatov:1976zz,Fadin:1975cb} and next-to-leading logarithmic (NLL) approximation~\cite{Fadin:1998py,Ciafaloni:1998gs} offers an effective framework to calculate the bulk of the cross sections. 

\begin{figure}[H]
 \centering
 \includegraphics[scale=0.5]{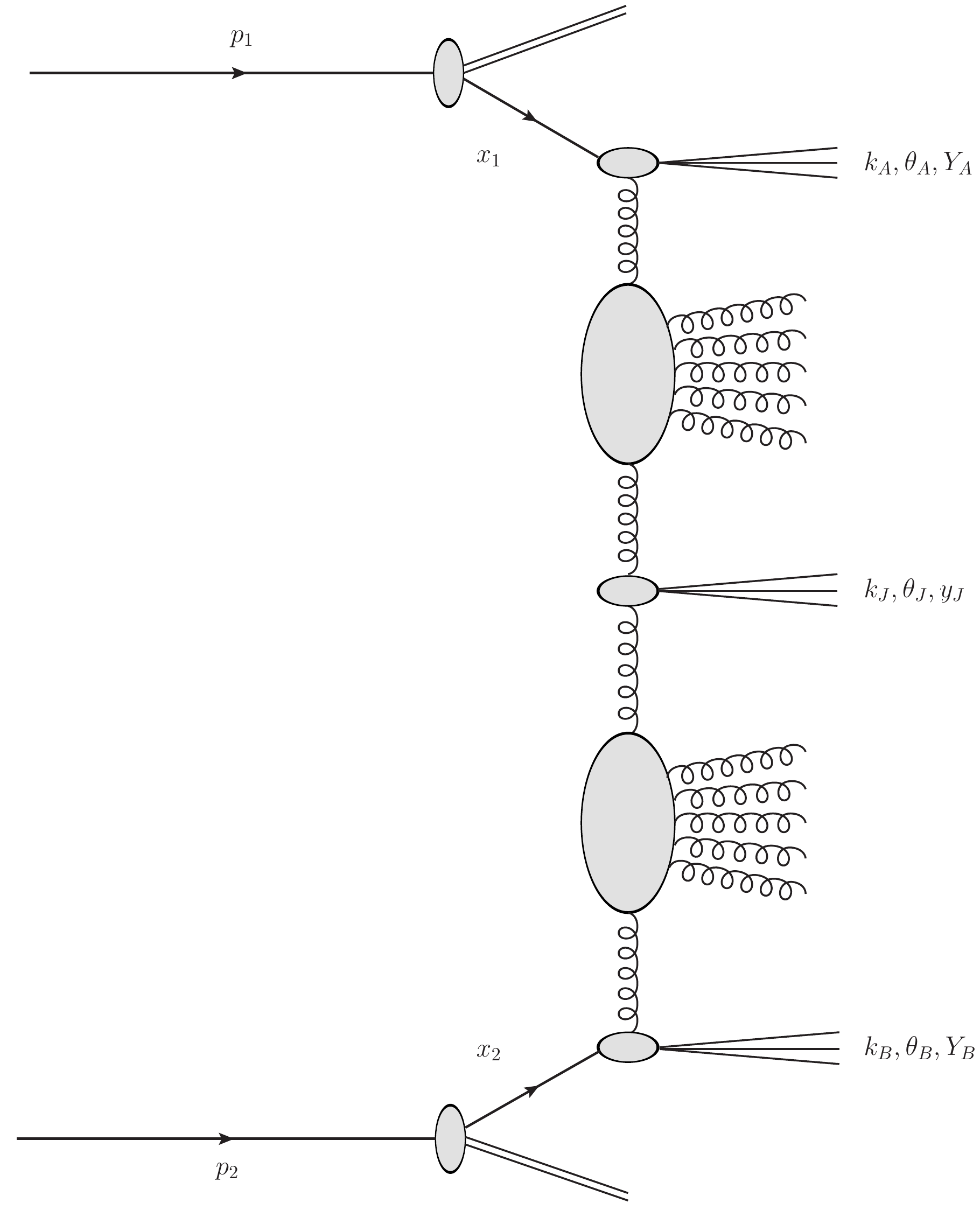}
 \caption[]
 {Inclusive three-jet production process in multi-Regge kinematics.}
 \label{fig:3j}
 \end{figure}Mueller-Navelet jets~\cite{Mueller:1986ey} correspond to the inclusive hadroproduction 
 of two jets\footnote{
 Another interesting idea,  suggested in~\cite{Ivanov:2012iv} 
and investigated in~\cite{Celiberto:2016hae}, is
the study of the production of two charged light hadrons,
$\pi^{\pm}$, $K^{\pm}$, $p$, $\bar p$, with large transverse momenta and well
separated in rapidity.} 
with large and similar transverse momenta, $k_{A,B}$,  and a significant relative separation in rapidity $Y=\ln ( x_1 x_2 s/(k_A k_B))$, where $x_{1,2}$ are the longitudinal momentum fractions of the partons generating the jets and $s$ is the centre-of-mass-energy squared $s$.
Different studies~\cite{DelDuca:1993mn,Stirling:1994he,Orr:1997im,Kwiecinski:2001nh,Angioni:2011wj,Caporale:2013uva, Caporale:2013sc,Marquet:2007xx,Colferai:2010wu,Ducloue:2013wmi,Ducloue:2014koa,Mueller:2015ael}  of the average values, $\langle \cos{(m \, \phi)} \rangle$, for the azimuthal-angle  formed by the two tagged jets, $\phi$, have shown the presence of a large soft gluon activity populating the rapidity gap. These observables are, however, strongly affected by collinear effects~\cite{Vera:2006un,Vera:2007kn}, stemming from the $n=0$ Fourier component in $\phi$ of the BFKL kernel. This dependence is removed if instead the ratios of projections on azimuthal-angle observables~${\cal R}^m_n = \langle \cos{(m \, \phi)} \rangle / \langle \cos{(n \, \phi)} \rangle$~\cite{Vera:2006un,Vera:2007kn} (where $m,n$ are integers and $\phi$ the azimuthal angle between the two tagged jets) are introduced. In particular, these also offer a more clear  signal of BFKL effects than the standard predictions for the growth of hadron structure functions $F_{2,L}$ (well fitted within  NLL approaches~\cite{Hentschinski:2012kr,Hentschinski:2013id}). The comparison of different NLL predictions for these ratios ${\cal R}^m_n$~\cite{Ducloue:2013bva,Caporale:2014gpa,Caporale:2014blm,Celiberto:2015dgl,Celiberto:2016ygs} with LHC experimental data has  been very successful. 

We understand the current situation as the beginning of precision physics using the BFKL formalism. Within the framework itself there exist many theoretical questions to be answered. Some of these include to find out what is the more accurate way to implement the running of the coupling, if there is any onset of saturation effects at the level of exclusive observables, how to isolate BFKL dynamics from multiple interaction effects, etc. To address these issues it is important to investigate even more exclusive final states.

 Here we advance in this direction by proposing new observables associated to the inclusive production of three jets: two of them are the original Mueller-Navelet jets and the third one is a tagged jet in central regions of rapidity (see Fig.~\ref{fig:3j}). Experimentally, they have the advantage to belong to the already recorded Mueller-Navelet events, it only requires of further binning in the internal jets. Theoretically, they will allow us to better understand distinct features of the BFKL ladder, in other words, to find out which ones of its predictions cannot be reproduced by other approaches such as low order exact perturbation theory or general-purpose Monte Carlo event generators. Parton-level studies have been recently presented in~\cite{Caporale:2015vya} while here we focus on calculating realistic cross-sections at the LHC. 

\begin{figure}[H]
 \centering
 \vspace{-.5cm}
 \includegraphics[scale=0.45]{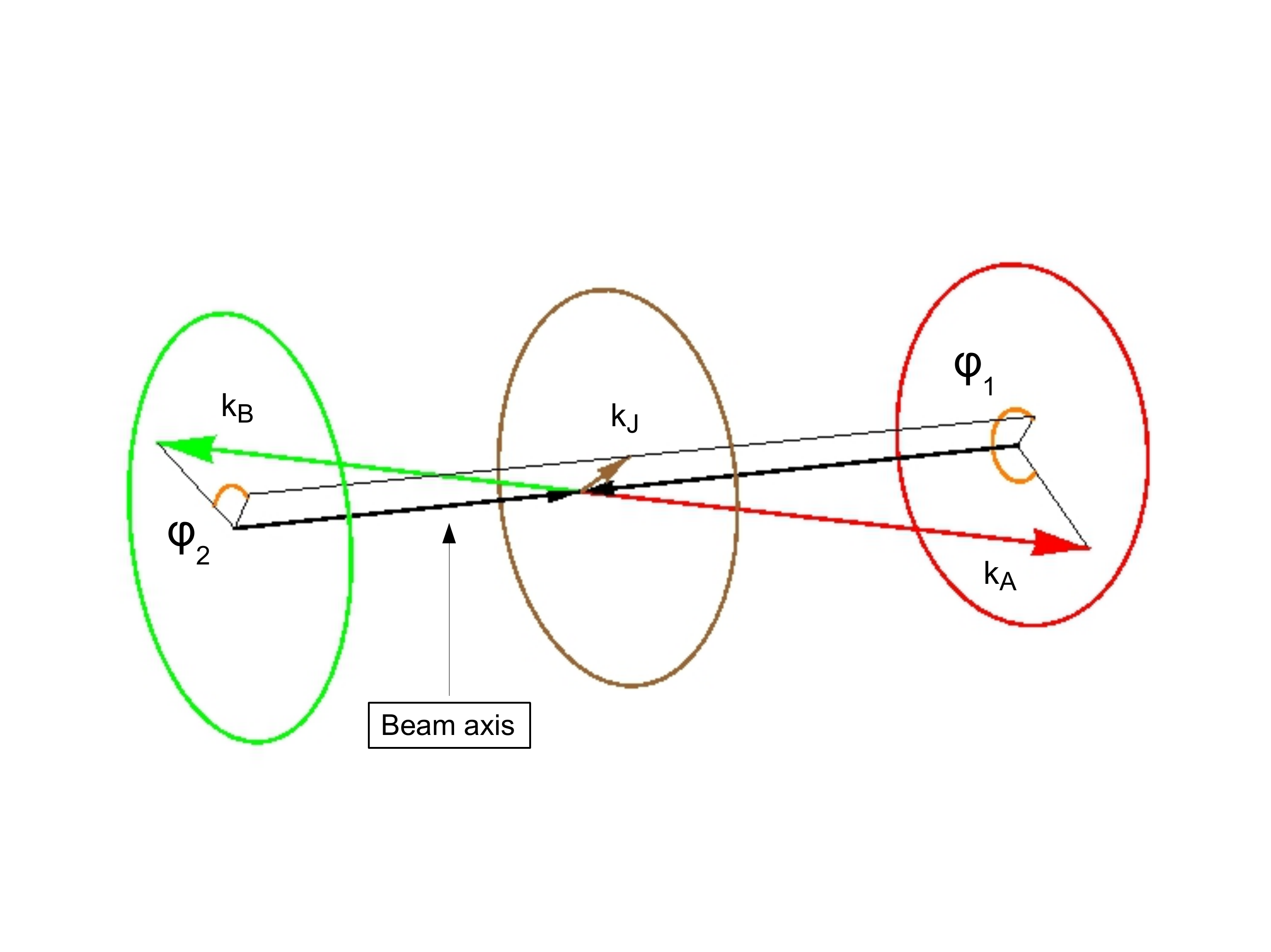}
 \vspace{-2cm}
 \caption{Representation of a three-jet event in a generic detector. All three
 circles are perpendicular to the beam axis.}
 \label{fig:3jdetector}
 \end{figure}In order to focus our discussion, we will present results for the above mentioned ${\cal R}^m_n$ ratios but now with a further dependence on the $p_t$ and rapidity of the central jet. As a novel result,  we will also present predictions for the new ratios 
\begin{eqnarray}
{\cal R}^{M N}_{P Q} =\frac{ \langle \cos{(M \, \phi_1)} \cos{(N \, \phi_2)} \rangle}{\langle \cos{(P \, \phi_1)} \cos{(Q \, \phi_2)} \rangle} \, , 
\label{Rmnpq}
\end{eqnarray}
where $\phi_1$ and $\phi_2$ are, respectively, the azimuthal angle difference between the first and the second (central) jet and between this one and the third jet (see Fig.~\ref{fig:3jdetector}). 

A further natural development in this direction has been the extension of these 
observables to the case of four-jet production in multi-Regge kinematics with a second tagged jet 
being produced in the central region of rapidity~\cite{Caporale:2015int,Caporale:2016xku}. 
This allows for the study of even more differential 
distributions in the transverse momenta, azimuthal angles and 
rapidities of the two central jets, for fixed values of the 
four momenta of the two forward (originally Mueller-Navelet) jets. The main observable 
${\cal R}^{M N L}_{P Q R}$ proposed at parton level in~\cite{Caporale:2015int} is the extension 
of the one in Eq.~(\ref{Rmnpq}), using three cosines instead of two in numerator and denominator.
This observable also paves the way for detailed studies of multiple parton scattering~\cite{Jung:2011yt,Baranov:2015nma,Maciula:2015vza,Maciula:2014pla,Kutak:2016mik} although we should point out that
in Ref.~\cite{Ducloue:2015jba} there is a claim that multiple parton
interactions (MPI) are negligible in the present LHC kinematics for the values of transverse momenta
used in the following.

In the next two Sections, we focus on the case of inclusive three-jet production performing a realistic study beyond the parton level calculation. This will allow for a comparison of our observables with forthcoming analysis of the LHC experimental data. Cross-sections are calculated using collinear factorization to produce the two most forward/backward jets, convoluting the ``hard" differential cross section, which follows the BFKL dynamics, with collinear parton distribution functions included in the forward ``jet vertex"~\cite{Caporale:2012:IF,Fadin:2000:gIF,Fadin:2000:qIF,Ciafaloni:1998kx,Ciafaloni:1998hu,Bartels:2001ge,Bartels:2002yj}. We link these two Mueller-Navelet jet-vertices with the centrally produced jet via two BFKL gluon Green functions. To simplify our predictions, we integrate over the momenta of all produced jets, using current LHC experimental cuts, only fixing the rapidity of the central jet to lie in the middle of the two most forward/backward tagged jets. In the following Section we will show the main formulas, in the next-to-last Section we will present our numerical predictions to finally end with our Summary and Outlook.

\section{Hadronic inclusive three-jet production in multi-Regge kinematics}

\begin{figure}[H]
 \centering
 \includegraphics[scale=0.6]{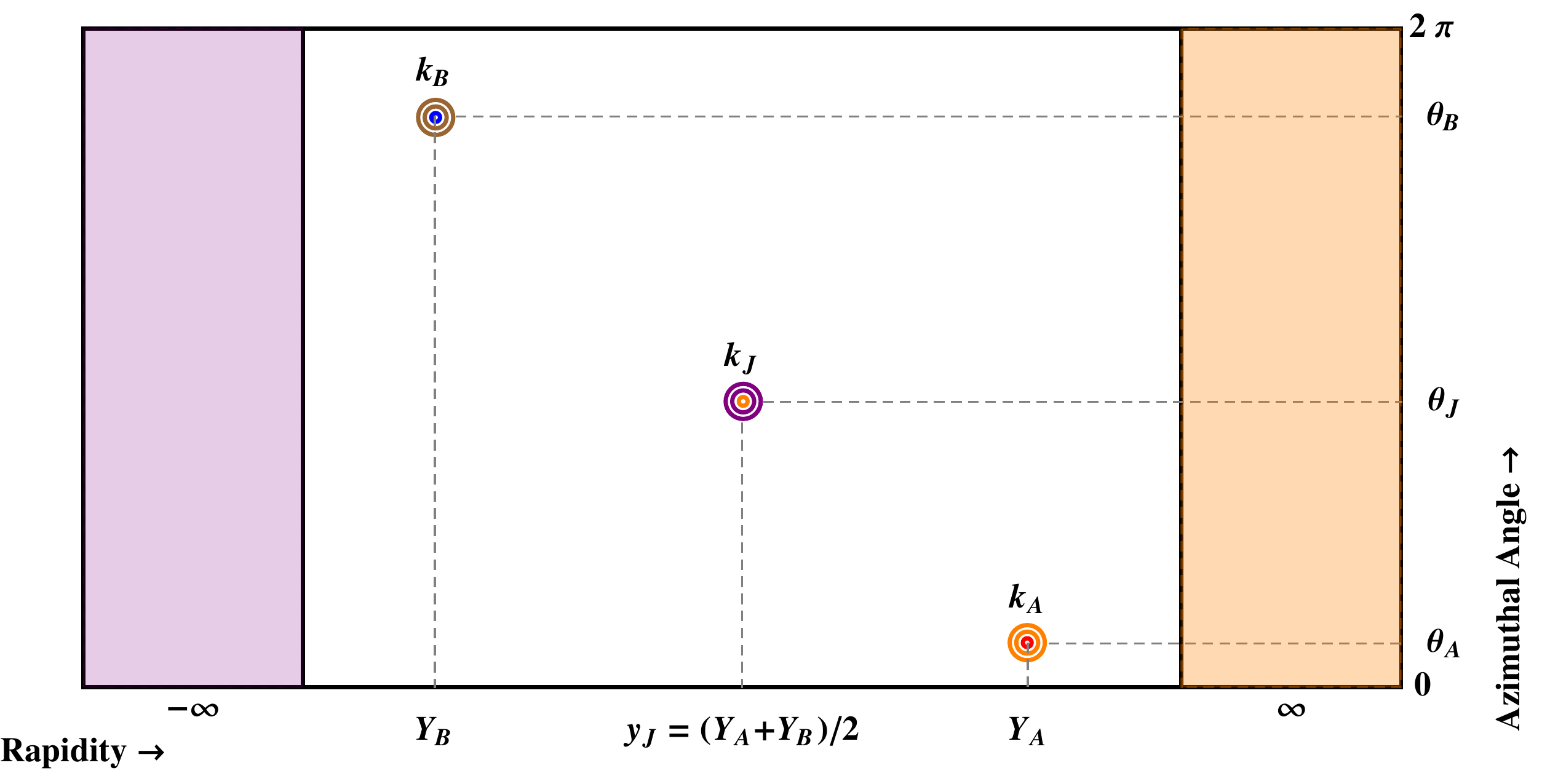}
 \caption[]
 {A primitive lego plot 
 depicting a three jet event. $k_A$ is a forward jet with 
 large positive
 rapidity $Y_A$ and azimuthal angle $\theta_A$, $k_B$ is a forward jet with 
 large negative
 rapidity $Y_B$ and azimuthal angle $\theta_B$ and
 $k_J$ is a central jet with 
 rapidity $y_J$ and azimuthal angle $\theta_J$. 
 }
 \label{fig:lego}
 \end{figure}

The process under investigation (see Figs.~\ref{fig:3j},~\ref{fig:3jdetector} and~\ref{fig:lego})
is the production of two forward/backward jets, both characterized by high transverse momenta $\vec{k}_{A,B}$ and well separated in rapidity, together with a third jet produced in the
central rapidity region and with possible associated mini-jet production. This corresponds to 
\begin{eqnarray}
\label{process}
{\rm proton }(p_1) + {\rm proton} (p_2) \to 
{\rm jet}(k_A) + {\rm jet}(k_J) + {\rm jet}(k_B)  + {\rm minijets}\;.
\end{eqnarray}

In collinear factorization the cross section for the process~(\ref{process}) reads
\begin{align}
\label{dsigma_pdf_convolution}
 & \frac{d\sigma^{3-{\rm jet}}}
      {dk_A \, dY_A \, d\theta_A \, 
       dk_B \, dY_B \, d\theta_B \, 
       dk_J \, dy_J d\theta_J}  = 
 \\ \nonumber 
 \hspace{1cm}& \sum_{r,s=q,{\bar q},g}\int_0^1 dx_1 \int_0^1 dx_2
 \ f_r\left(x_1,\mu_F\right)
 \ f_s\left(x_2,\mu_F\right) \;
 d{\hat\sigma}_{r,s}\left(\hat{s},\mu_F\right) \;,
\end{align}
where the $r, s$ indices specify the parton types 
(quarks $q = u, d, s, c, b$;
antiquarks $\bar q = \bar u, \bar d, \bar s, \bar c, \bar b$; or gluon $g$),
$f_{r,s}\left(x, \mu_F \right)$ are the initial proton PDFs; 
$x_{1,2}$ represent the longitudinal fractions of the partons involved 
in the hard subprocess; $d\hat\sigma_{r,s}\left(\hat{s}, \mu_F \right)$ 
is the partonic cross section for the production of jets and
$\hat{s} \equiv x_1x_2s$ is the squared center-of-mass energy of the
hard subprocess (see Fig.~\ref{fig:3j}). The BFKL dynamics enters in the cross-section 
for the partonic hard subprocess $d{\hat\sigma}_{r,s}$ in the form of two forward gluon Green functions $\varphi$ to be described below. 

Using the definition of the jet vertex in the leading order approximation~\cite{Caporale:2012:IF}, 
we can present the cross section for the process as
\begin{align}
 & \frac{d\sigma^{3-{\rm jet}}}
      {dk_A \, dY_A \, d\theta_A \, 
       dk_B \, dY_B \, d\theta_B \, 
       dk_J \, dy_J d\theta_J} = 
 \nonumber \\ \hspace{1cm}&  
 \frac{8 \pi^3 \, C_F \, \asb^3}{N_C^3} \, 
 \frac{x_{J_A} \, x_{J_B}}{k_A \, k_B \, k_J} \,
 \int d^2 \vec{p}_A \int d^2 \vec{p}_B \,
 \delta^{(2)} \left(\vec{p}_A + \vec{k}_J- \vec{p}_B\right) \,
 \nonumber \\  \hspace{1cm}& \times 
 \left(\frac{N_C}{C_F}f_g(x_{J_A},\mu_F)
 +\sum_{r=q,\bar q}f_r(x_{J_A},\mu_F)\right) \,
 \nonumber \\ \hspace{1cm}& \times
 \left(\frac{N_C}{C_F}f_g(x_{J_B},\mu_F)
 +\sum_{s=q,\bar q}f_s(x_{J_B},\mu_F)\right)
 \nonumber \\ \hspace{1cm}& \times
 \varphi \left(\vec{k}_A,\vec{p}_A,Y_A - y_J\right) 
 \varphi \left(\vec{p}_B,\vec{k}_B,y_J - Y_B\right).
\end{align}
In order to lie within multi-Regge kinematics, we have considered the ordering in the rapidity of the produced particles $Y_A > y_J > Y_B$, while $k_J^2$ is always 
above the experimental resolution scale.
$x_{J_{A,B}}$ are the longitudinal momentum fractions
of the two external jets, linked to the respective rapidities 
$Y_{J_{A,B}}$ by the relation 
$x_{J_{A,B}} = k_{A,B} \, e^{\, \pm \, Y_{J_{A,B}}} / \sqrt{s}$. 
$\varphi$ are BFKL gluon Green functions normalized to 
$ \varphi \left(\vec{p},\vec{q},0\right) = \delta^{(2)} \left(\vec{p} - \vec{q}\right)$ 
and $\bar{\alpha}_s =  N_c/\pi \, \alpha_s \left(\mu_R\right)$.

Building up on the work in Ref.~\cite{Caporale:2015vya,Caporale:2015int}, 
we study observables for which the BFKL approach will be distinct from other formalisms and also 
rather insensitive to possible higher order corrections. We focus on new quantities 
whose associated distributions are different from the ones which
characterize the Mueller-Navelet case, though still related 
to the azimuthal-angle correlations by projecting the differential cross section
on the two relative azimuthal angles between each external jet
and the central one 
$\Delta\theta_{\widehat{AJ}} = \theta_A - \theta_J - \pi$ and 
$\Delta\theta_{\widehat{JB}} = \theta_J - \theta_B - \pi$ (see Fig.~\ref{fig:lego}). 
Taking into account the factors coming from the jet vertices, 
it is possible to rewrite Eq.~(7) of~\cite{Caporale:2015vya} 
in the form
\begin{align}
 & \int_0^{2 \pi} d \theta_A \int_0^{2 \pi} d \theta_B \int_0^{2 \pi} 
 d \theta_J \cos{\left(M \Delta\theta_{\widehat{AJ}} \right)} \,
            \cos{\left(N \Delta\theta_{\widehat{JB}} \right)}
 \nonumber \\ 
 & \hspace{0.5cm} 
 \frac{d\sigma^{3-{\rm jet}}}
      {dk_A \, dY_A \, d\theta_A \, 
       dk_B \, dY_B \, d\theta_B \, 
       dk_J \, dy_J d\theta_J}    = 
 \nonumber \\
 &  \hspace{0.07cm} 
 \frac{8 \pi^4 \, C_F \, \asb^3}{N_C^3} \, 
 \frac{x_{J_A} \, x_{J_B}}{k_A \, k_B} 
 \left(\frac{N_C}{C_F}f_g(x_{J_A},\mu_F) \,
 +\sum_{r=q,\bar q}f_r(x_{J_A},\mu_F)\right) \,
 \nonumber \\ 
 & \times \hspace{0.07cm}
 \left(\frac{N_C}{C_F}f_g(x_{J_B},\mu_F)
 +\sum_{s=q,\bar q}f_s(x_{J_B},\mu_F)\right) \, 
 \sum_{L=0}^{N} 
 \left( \begin{array}{c}
 \hspace{-.2cm}N \\
 \hspace{-.2cm}L\end{array} \hspace{-.18cm}\right)
 \left(k_J^2\right)^{\frac{L-1}{2}}
 \nonumber \\ 
 & \times \hspace{0.07cm}
 \int_{0}^\infty d p^2 \, \left(p^2\right)^{\frac{N-L}{2}} \,
 \int_0^{2 \pi}  d \theta    \frac
 {(-1)^{M+N} \cos{ \left(M \theta\right)} \cos{\left((N-L) \theta\right)}}
 {\sqrt{\left(p^2 + k_J^2+ 2 p k_J \cos{\theta}\right)^{N}}}
 \nonumber \\ 
 & \times \hspace{0.07cm}
 \varphi_{M} \left(k_A^2,p^2,Y_A-y_J\right)
 \varphi_{N} \left(p^2+ k_J^2 + 2 p k_J \cos{\theta},
                k_B^2,y_J-Y_B\right),
\end{align}
where 
\begin{align}
 \varphi_{n} \left(k^2,q^2,y\right) \; &= \; 
 2 \, \int_0^\infty d \nu   
 \cos{\left(\nu \ln{\frac{k^2}{q^2}}\right)}  
 \frac{e^{\bar{\alpha}_s  \chi_{|n|} \left(\nu\right) y}}
      {\pi \sqrt{k^2 q^2} }, \label{phin}
 \\
 \chi_{n} \left(\nu\right) \; &= \; 2\, \psi (1) - 
 \psi \left( \frac{1+n}{2} + i \nu\right) - 
 \psi \left(\frac{1+n}{2} - i \nu\right)
\end{align}
($\psi$ is the logarithmic derivative of Euler's gamma function).

The related experimental observable we propose corresponds to
the mean value (with $M,N$ being positive integers)
\begin{eqnarray}
\label{Cmn}
 {\cal C}_{MN} \, = \,
 \langle \cos{\left(M \left( \theta_A - \theta_J - \pi\right)\right)}  
 \cos{\left(N \left( \theta_J - \theta_B - \pi\right)\right)}
 \rangle && \\
 &&\hspace{-9cm} = \frac{\int_0^{2 \pi} d \theta_A d \theta_B d \theta_J \cos{\left(M \left( \theta_A - \theta_J - \pi\right)\right)}  \cos{\left(N \left( \theta_J - \theta_B - \pi\right)\right)}
 d\sigma^{3-{\rm jet}} }{\int_0^{2 \pi} d \theta_A d \theta_B d \theta_J 
 d\sigma^{3-{\rm jet}} }.\nonumber
\end{eqnarray}

From a phenomenological point of view, since our main target is to provide testable predictions compatible with
the current and future experimental data, we now introduce those kinematical cuts already in place at the LHC. For this purpose, we integrate ${\cal C}_{M,N}$ over the momenta of the tagged jets in the form
\begin{align}
\label{Cmn_int}
 & 
 C_{MN} =
 \nonumber \\
 &
 \int_{Y_A^{\rm min}}^{Y_A^{\rm max}} \hspace{-0.25cm} dY_A
 \int_{Y_B^{\rm min}}^{Y_B^{\rm max}} \hspace{-0.25cm} dY_B
 \int_{k_A^{\rm min}}^{k_A^{\rm max}} \hspace{-0.25cm} dk_A
 \int_{k_B^{\rm min}}^{k_B^{\rm max}} \hspace{-0.25cm} dk_B
 \int_{k_J^{\rm min}}^{k_J^{\rm max}} \hspace{-0.25cm} dk_J
 \delta\left(Y_A - Y_B - Y\right) {\cal C}_{MN},
\end{align}
where the forward/backward jet rapidities are taken in the
range delimited by $Y_A^{\rm min} = Y_B^{\rm min} = -4.7$  and 
$Y_A^{\rm max} = Y_B^{\rm max} = 4.7$, keeping their difference 
$Y \equiv Y_A - Y_B$ fixed at definite values in the range $5 < Y < 9$.

From a more theoretical perspective, it is important to have as good as possible perturbative 
stability  in our 
predictions (see~\cite{Caporale:2013uva} for a related discussion). 
This can be achieved by removing the contribution stemming  
from the zero conformal spin, 
which corresponds to the index $n=0$ in Eq.~(\ref{phin}).
We, therefore, introduce the ratios
\begin{eqnarray}
\label{RPQMN}
R_{PQ}^{MN} \, = \, \frac{C_{MN}}{C_{PQ}}
\label{RmnqpNew}
\end{eqnarray}
which are free from any $n=0$ dependence. We proceed now to present our numerical results for a number of different kinematic configurations.

\section{Numerical results for azimuthal-angle dependences}

We now study the ratios  $R_{PQ}^{MN}(Y)$ in Eq.~(\ref{RmnqpNew}) as functions of the 
rapidity difference Y between the most forward and the most backward jets 
for a set of characteristic values of $M, N, P, Q$ and for two different 
center-of-mass energies: $\sqrt s = 7$ and $\sqrt s = 13$ TeV. Since we are integrating over $k_A$ and $k_B$,  
we have the opportunity to impose either symmetric or asymmetric
cuts, as it has been previously done in the  Mueller-Navelet case~\cite{Ducloue:2013wmi,Celiberto:2015dgl}.
To be more precise, we study the two kinematical configurations:
\begin{enumerate}
\item $k_A^{\rm min} = 35$ GeV, $k_B^{\rm min} = 35$ GeV, $k_A^{\rm max} = k_B^{\rm max}  = 60$ GeV
(symmetric); \,
\item $k_A^{\rm min} = 35$ GeV, $k_B^{\rm min} = 50$ GeV,  $k_A^{\rm max} = k_B^{\rm max}  = 60$ GeV
(asymmetric).
\end{enumerate}

In order to be as close as possible to the rapidity ordering characteristic of multi-Regge kinematics, 
we set the value of the central jet rapidity 
such that it is equidistant to $Y_A$ and $Y_B$ by imposing
the condition $y_J = \frac{Y_A + Y_B}{2}$. Moreover, since by tagging
a central jet we are able to extract more exclusive information from our
observables, we allow three possibilities for the transverse momentum
$k_J$, that is, $20\, \mathrm{GeV} < k_J < 35\, \mathrm{GeV}$ (bin-1),
$35 \,\mathrm{GeV} < k_J < 60\, \mathrm{GeV}$ (bin-2) and
$60\, \mathrm{GeV} < k_J < 120\, \mathrm{GeV}$ (bin-3). Keeping in mind that 
the forward/backward jets have transverse momenta in the  range
$\left[35 \,\mathrm{GeV}, 60 \,\mathrm{GeV}\right]$, restricting the value
of $k_J$ within these three bins allows us to see how the ratio
$R_{PQ}^{MN}(Y)$ changes behaviour depending on the relative size of the
central jet when compared to the forward/backward ones.
Bin-1, bin-2 and bin-3 correspond to $k_J$ being smaller than, similar to and larger than
$k_A$, $k_B$, respectively.

Before we proceed to present our numerical results, we should note that
we performed the numerical computation of the ratios 
$\mathcal{R}^{MN}_{PQ}$ both 
in \textsc{Fortran} and in \textsc{Mathematica} (mainly for cross-checks).
The NLO MSTW 2008 PDF sets~\cite{MSTW:2009} were used 
and for the strong coupling $\alpha_s$ we chose 
a two-loop running coupling setup 
with $\alpha_s\left(M_Z\right)=0.11707$. 
We made extensive use of the integration routine 
\cod{Vegas}~\cite{VegasLepage:1978} 
as implemented in the \cod{Cuba} library~\cite{Cuba:2005,ConcCuba:2015}.
Furthermore, we used the \cod{Quadpack} library~\cite{Quadpack:book:1983}
and a slightly modified version 
of the \cod{Psi}~\cite{RpsiCody:1973} routine.

\begin{figure}[p]
\newgeometry{left=-10cm,right=1cm}
\vspace{-2cm}

   \hspace{-2.25cm}
   \includegraphics[scale=0.45]{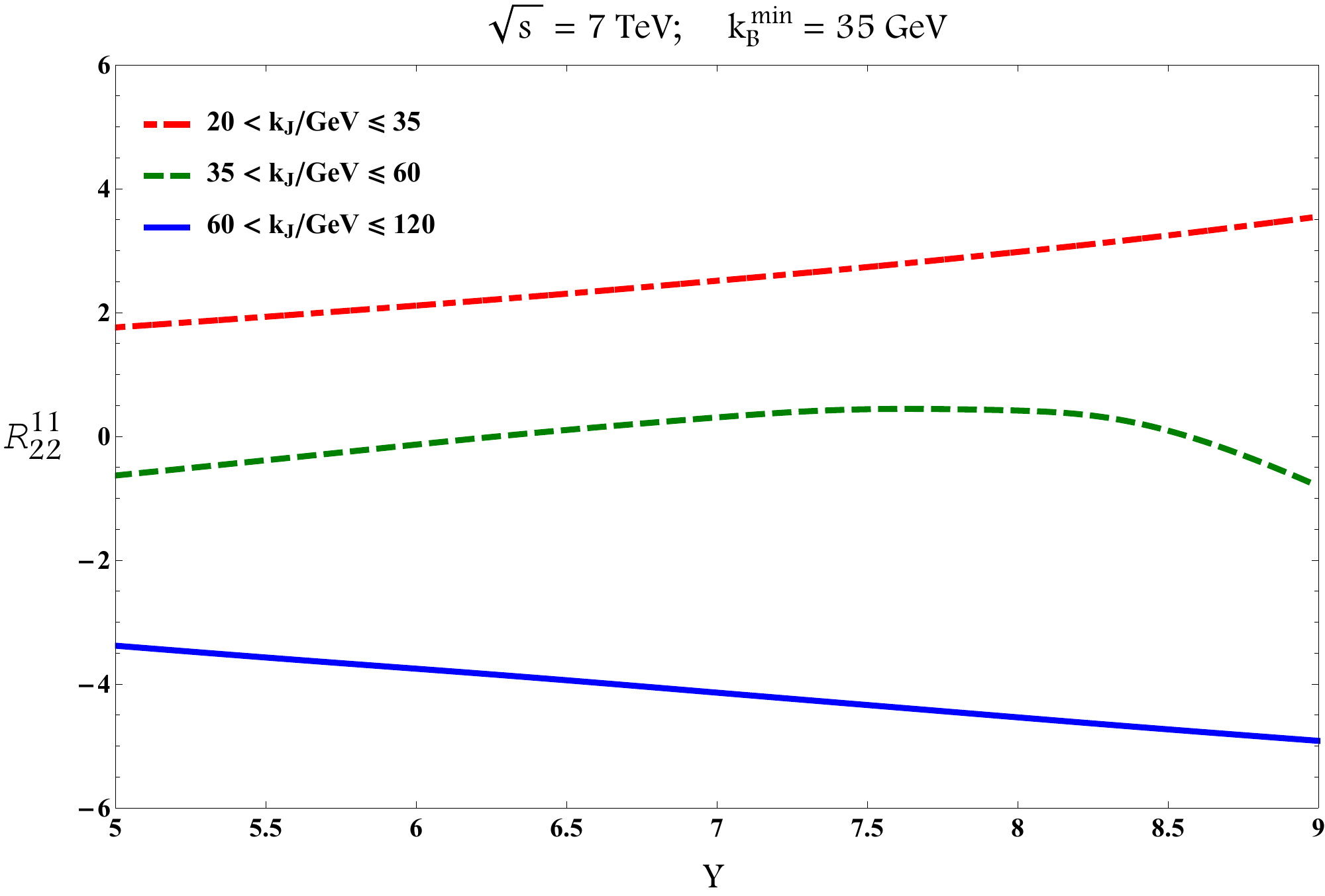}
   \includegraphics[scale=0.45]{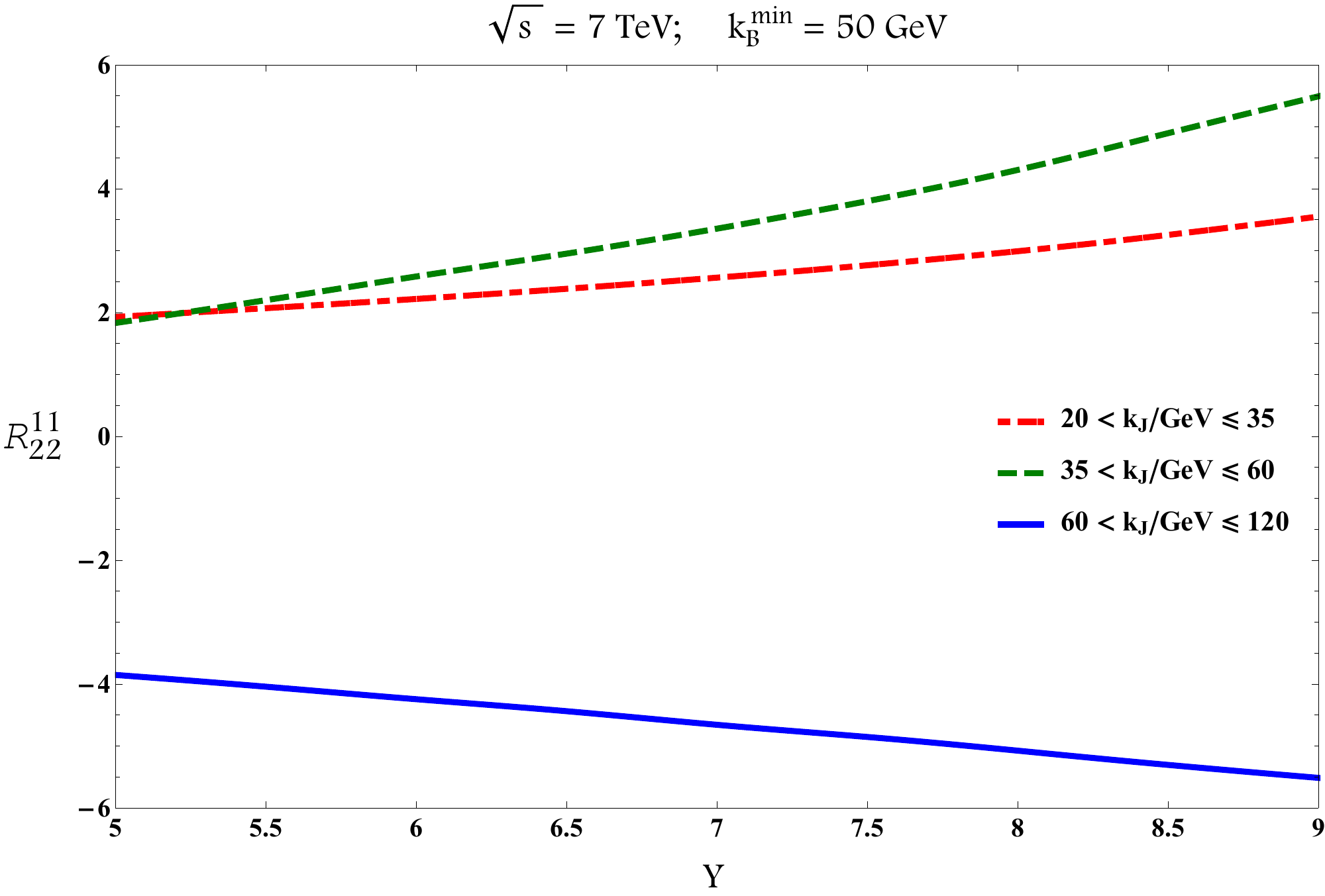}
   \vspace{1cm}

   \hspace{-2.25cm}
   \includegraphics[scale=0.45]{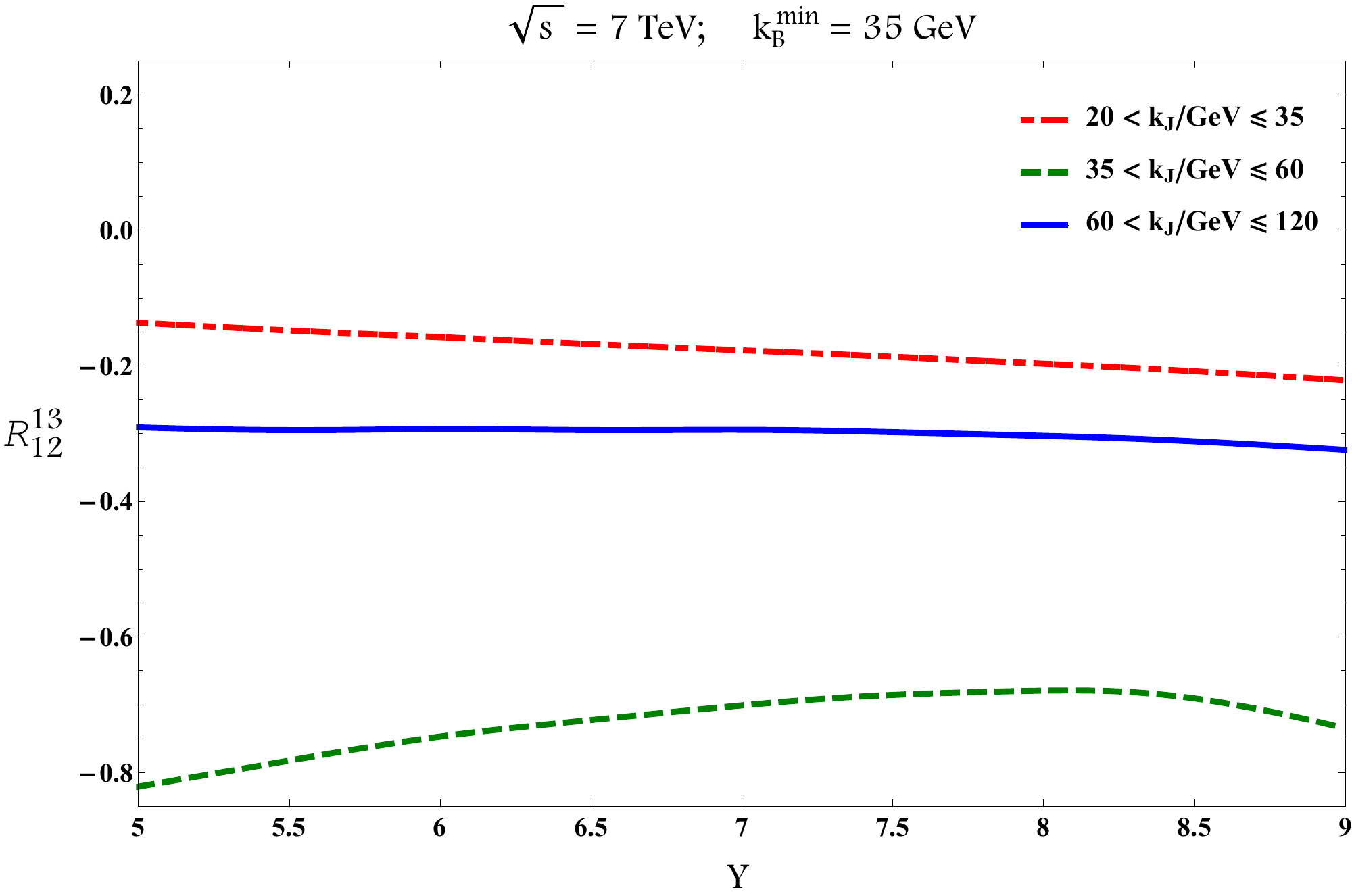}
   \includegraphics[scale=0.45]{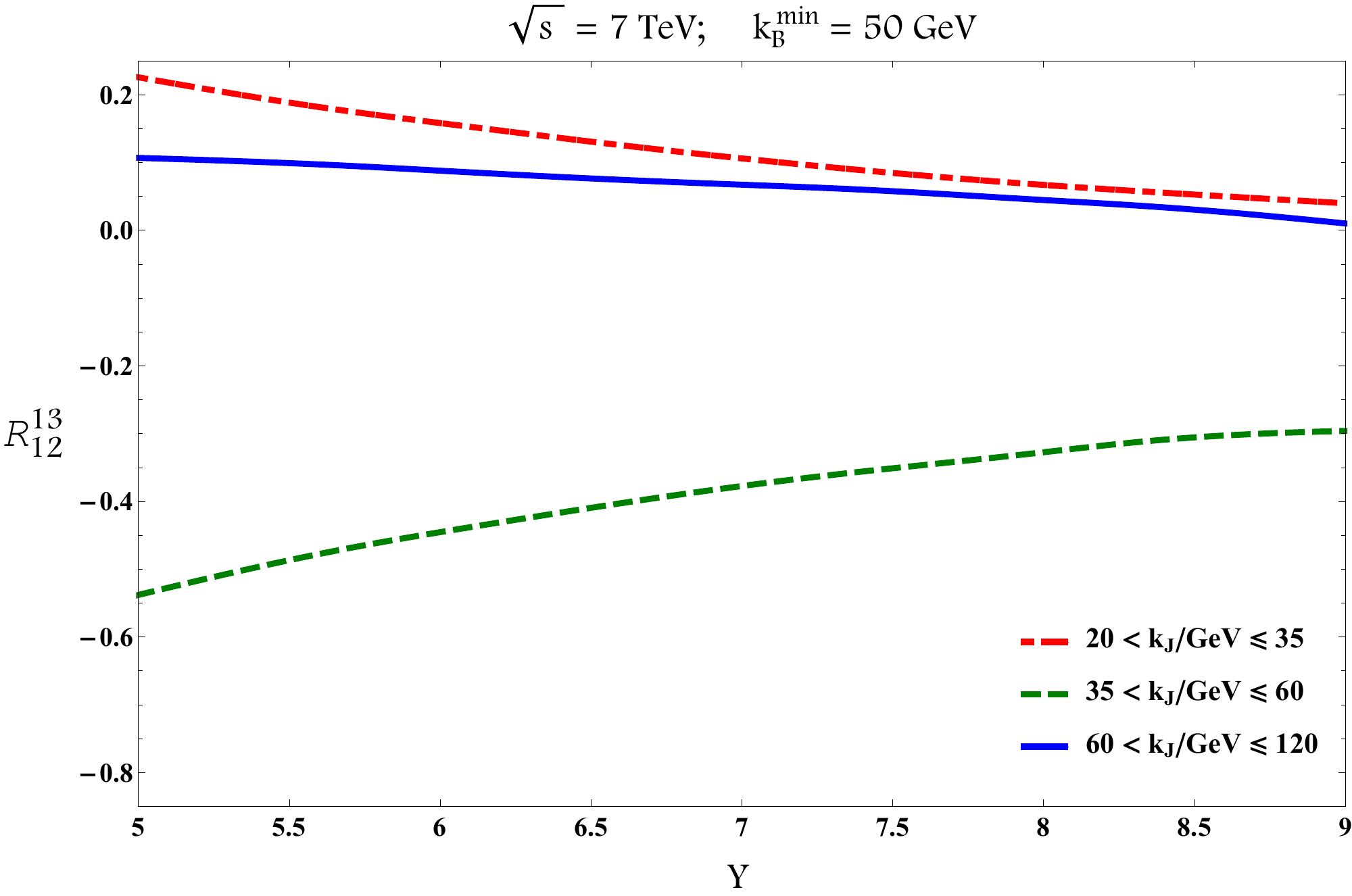}
   \vspace{1cm}

   \hspace{-2.25cm}   
   \includegraphics[scale=0.45]{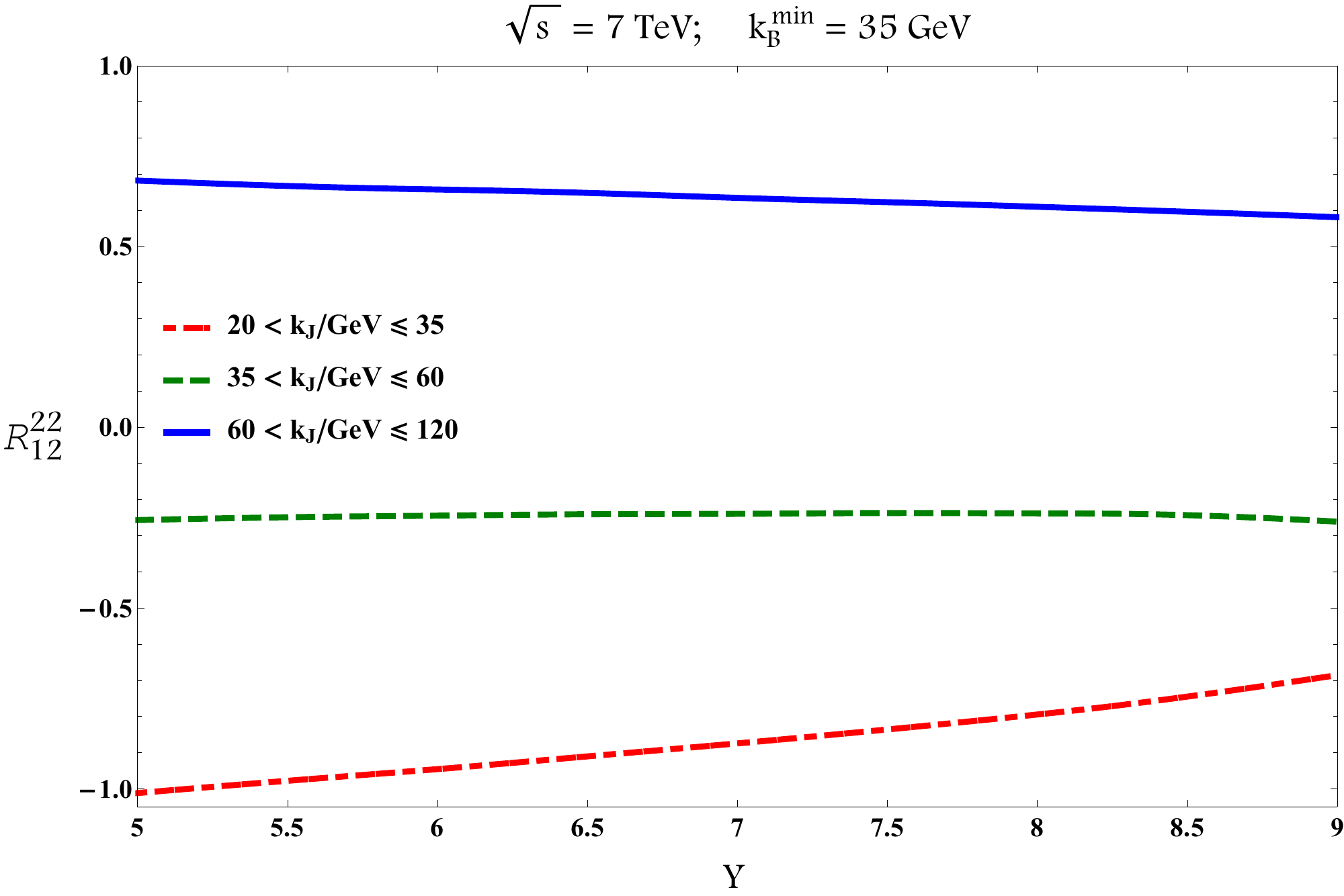}
   \includegraphics[scale=0.45]{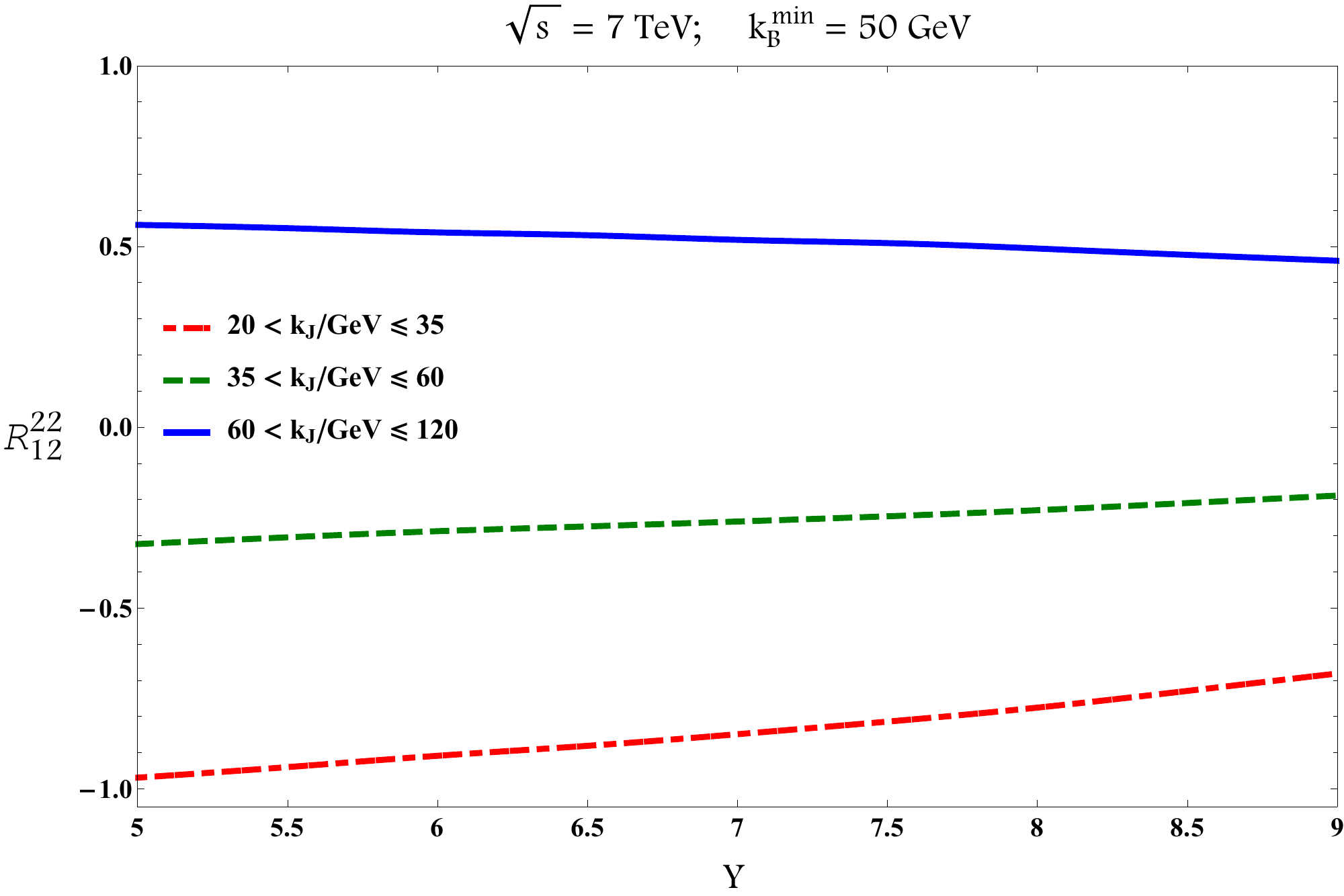}

\restoregeometry
\caption{\small $Y$-dependence of 
$R^{11}_{22}$, $R^{13}_{12}$ and $R^{22}_{12}$ 
for $\sqrt s = 7$ TeV and $k_B^{\rm min} = 35$ GeV (left column)
and $k_B^{\rm min} = 50$ GeV (right column).} 
\label{fig:7-first}
\end{figure}

\begin{figure}[p]
\newgeometry{left=-10cm,right=1cm}
\vspace{-2cm}

   \hspace{-2.25cm}
   \includegraphics[scale=0.45]{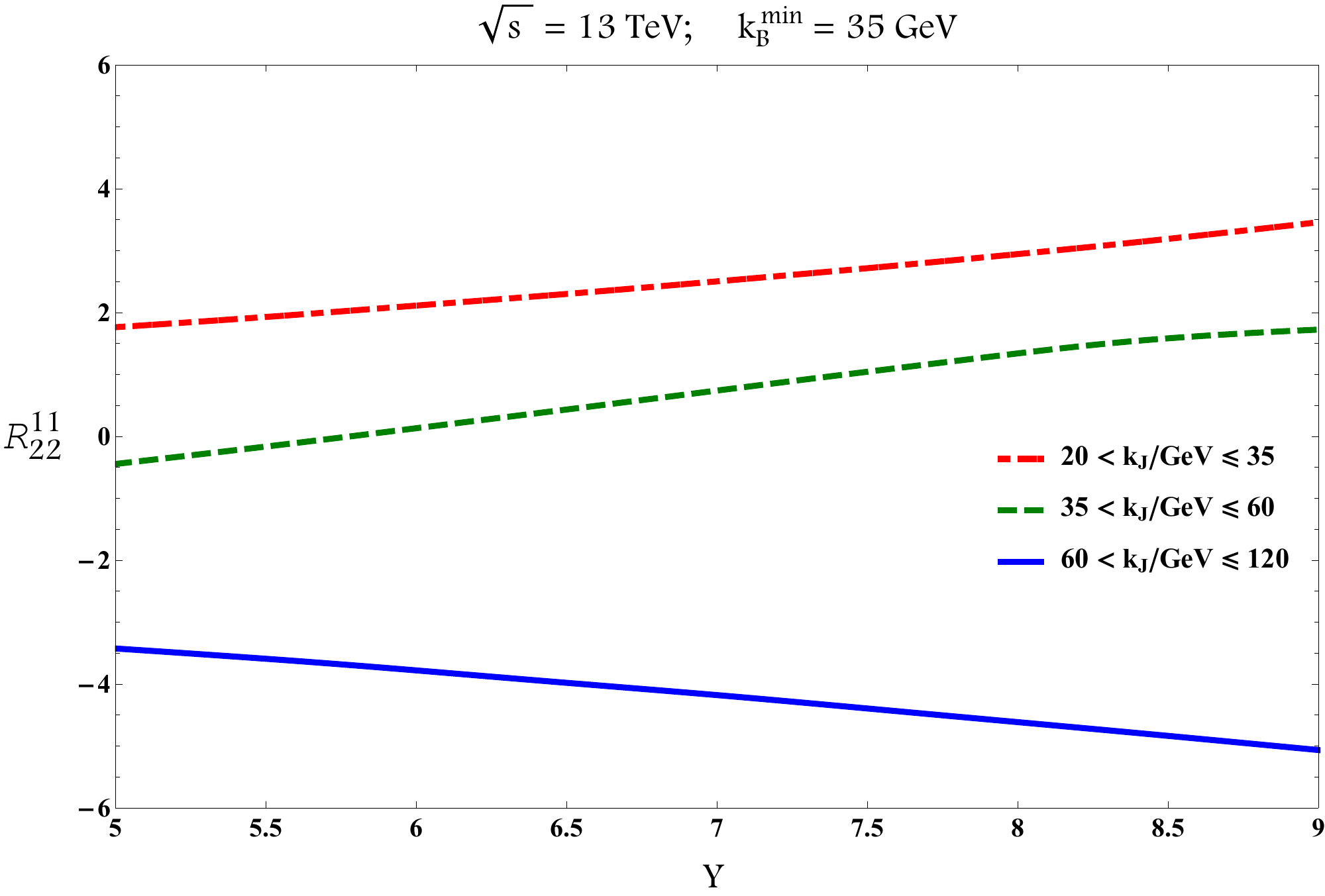}
   \includegraphics[scale=0.45]{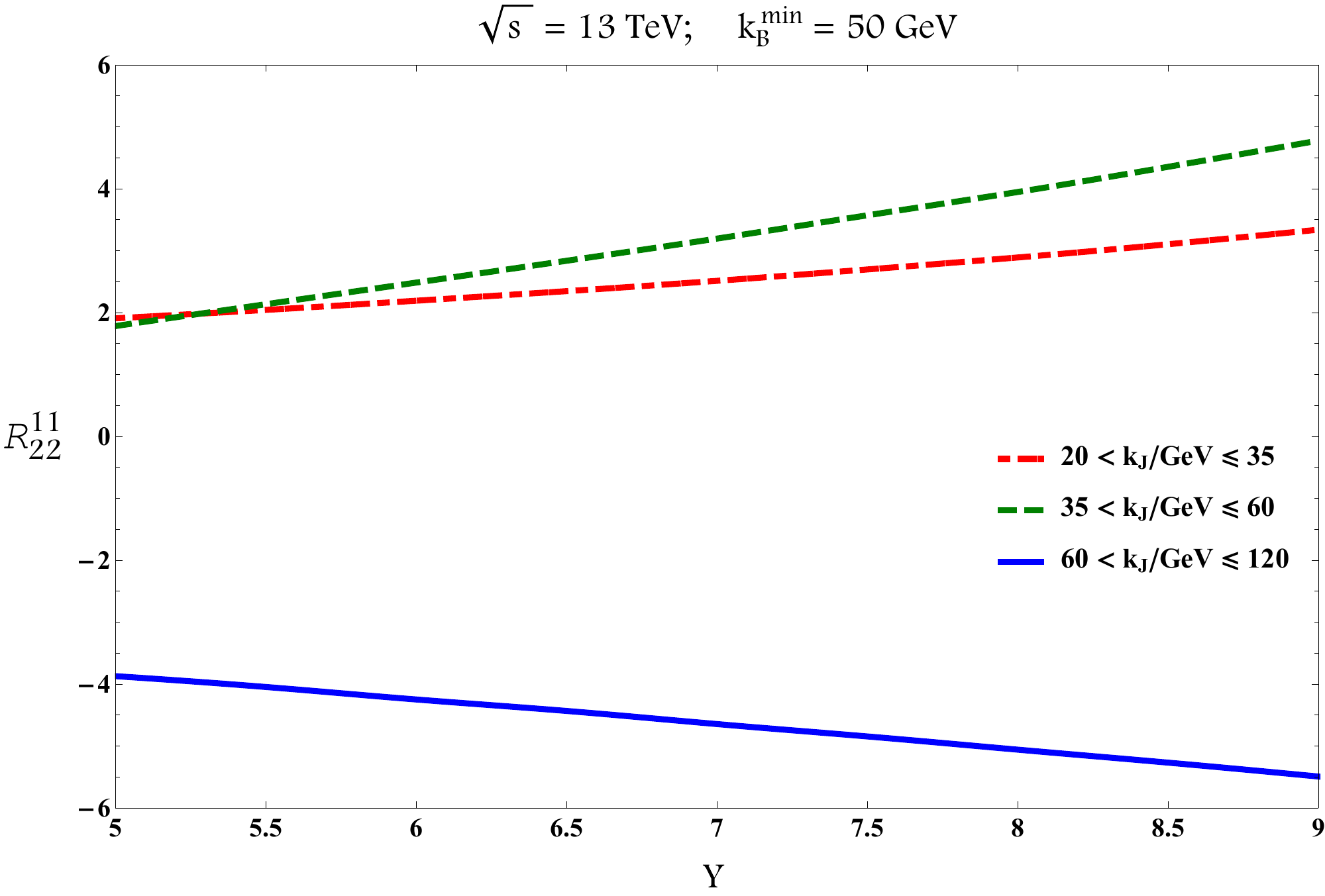}
   \vspace{1cm}

   \hspace{-2.25cm}
   \includegraphics[scale=0.45]{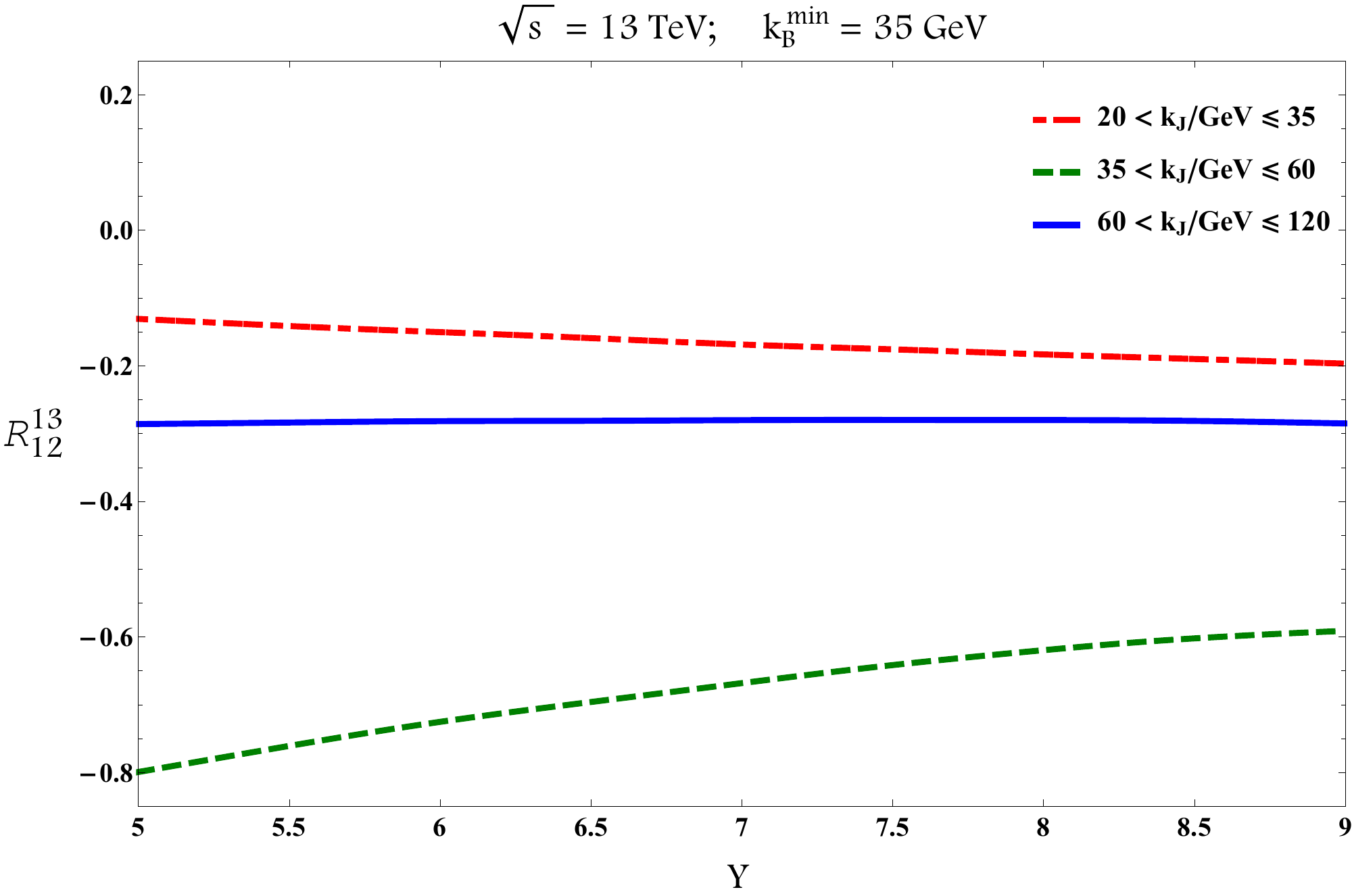}
   \includegraphics[scale=0.45]{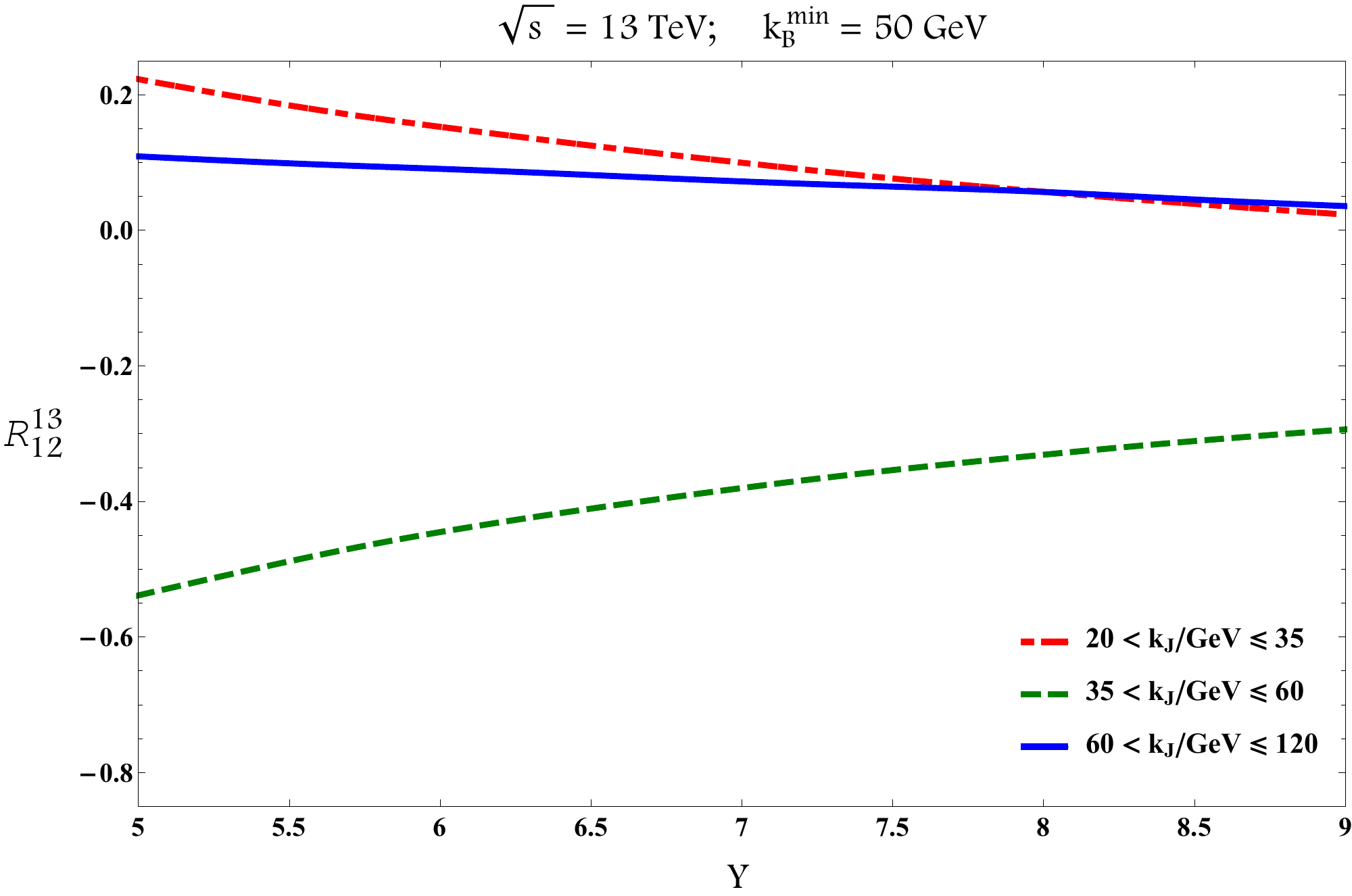}
   \vspace{1cm}

   \hspace{-2.25cm}   
   \includegraphics[scale=0.45]{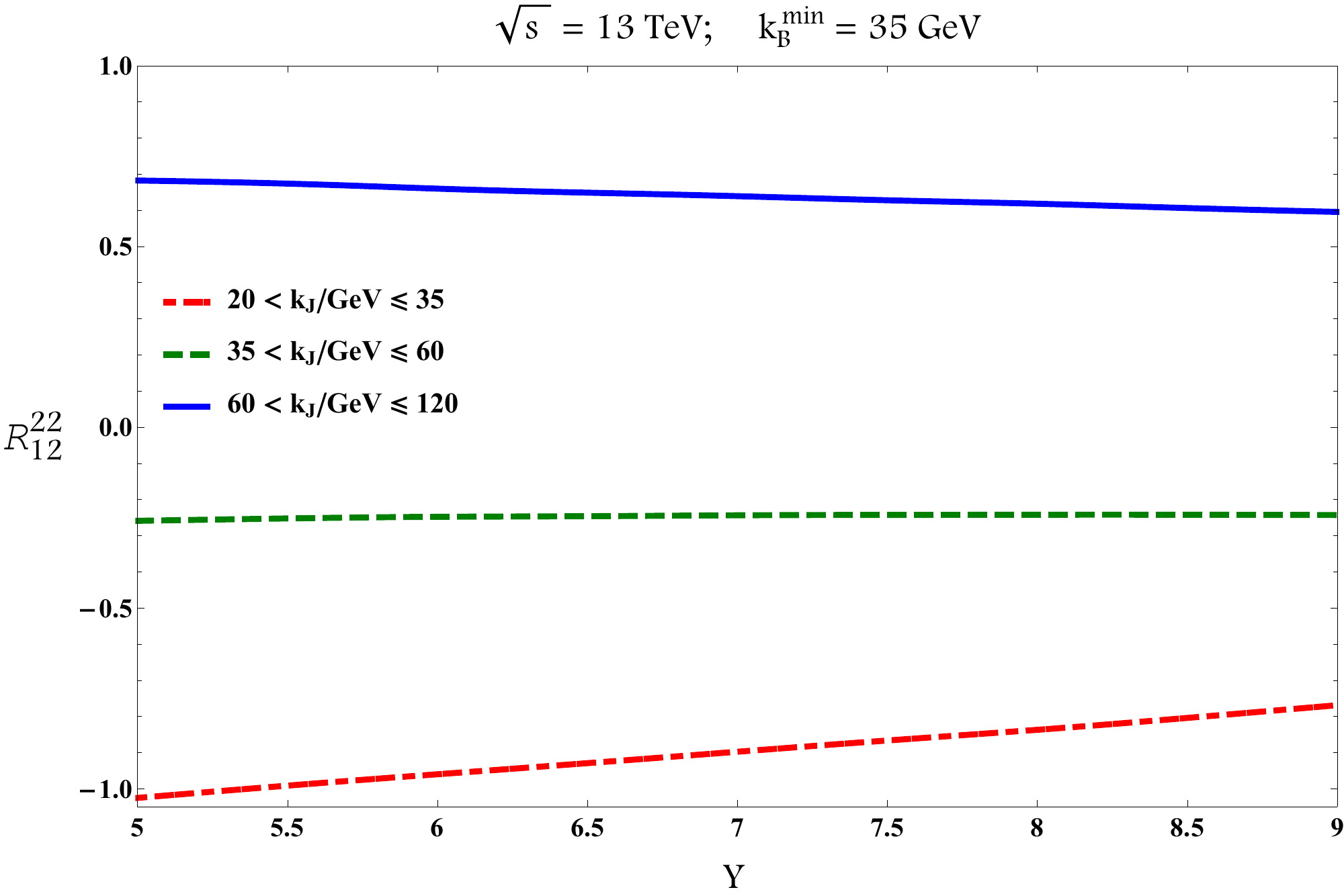}
   \includegraphics[scale=0.45]{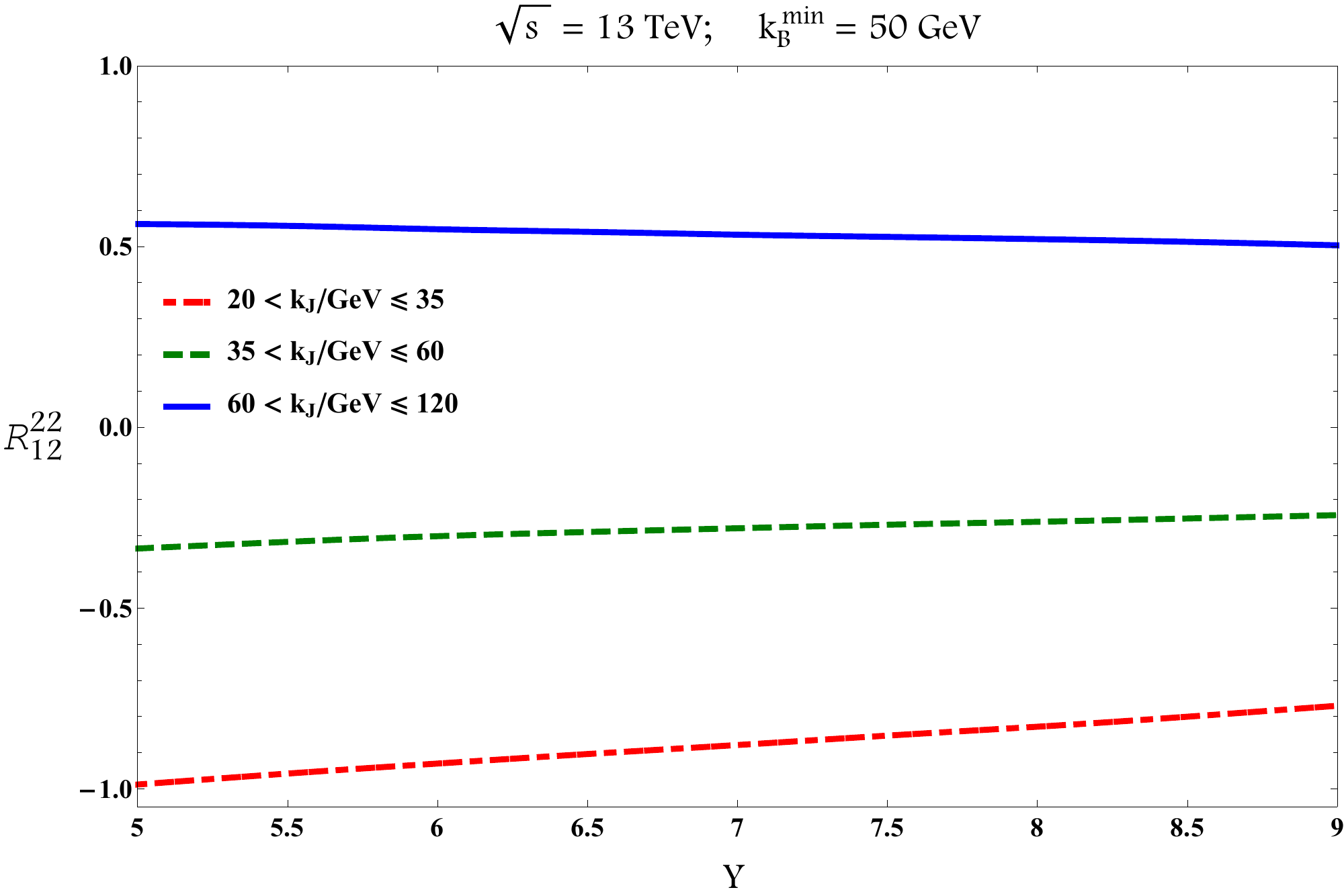}

\restoregeometry
\caption{\small $Y$-dependence of 
$R^{11}_{22}$, $R^{13}_{12}$ and $R^{22}_{12}$ 
for $\sqrt s = 13$ TeV and $k_B^{\rm min} = 35$ GeV (left column)
and $k_B^{\rm min} = 50$ GeV (right column).} 
\label{fig:13-first}
\end{figure}

\begin{figure}[p]
\newgeometry{left=-10cm,right=1cm}
\vspace{-2cm}

   \hspace{-2.25cm}
   \includegraphics[scale=0.45]{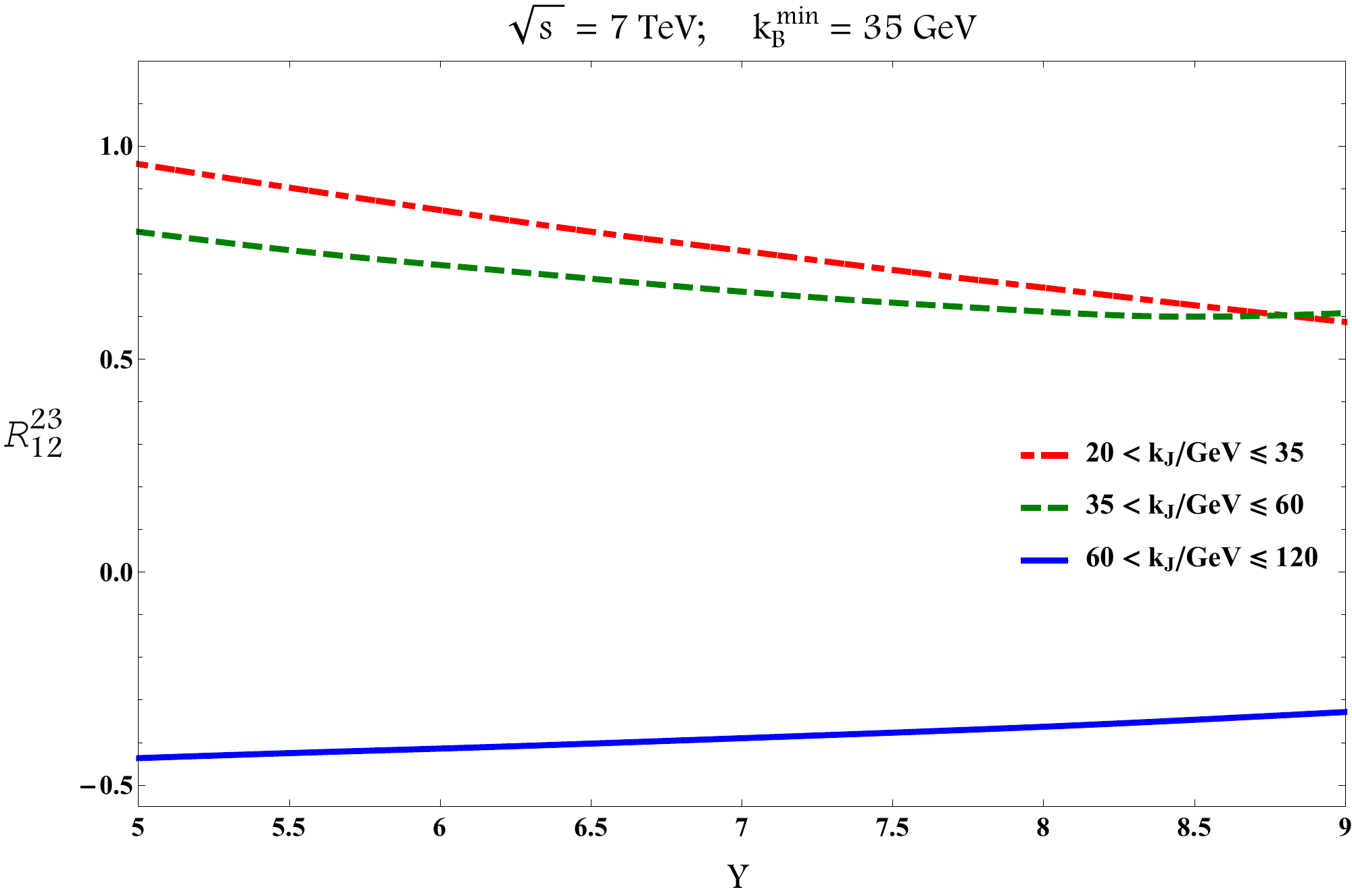}
   \includegraphics[scale=0.45]{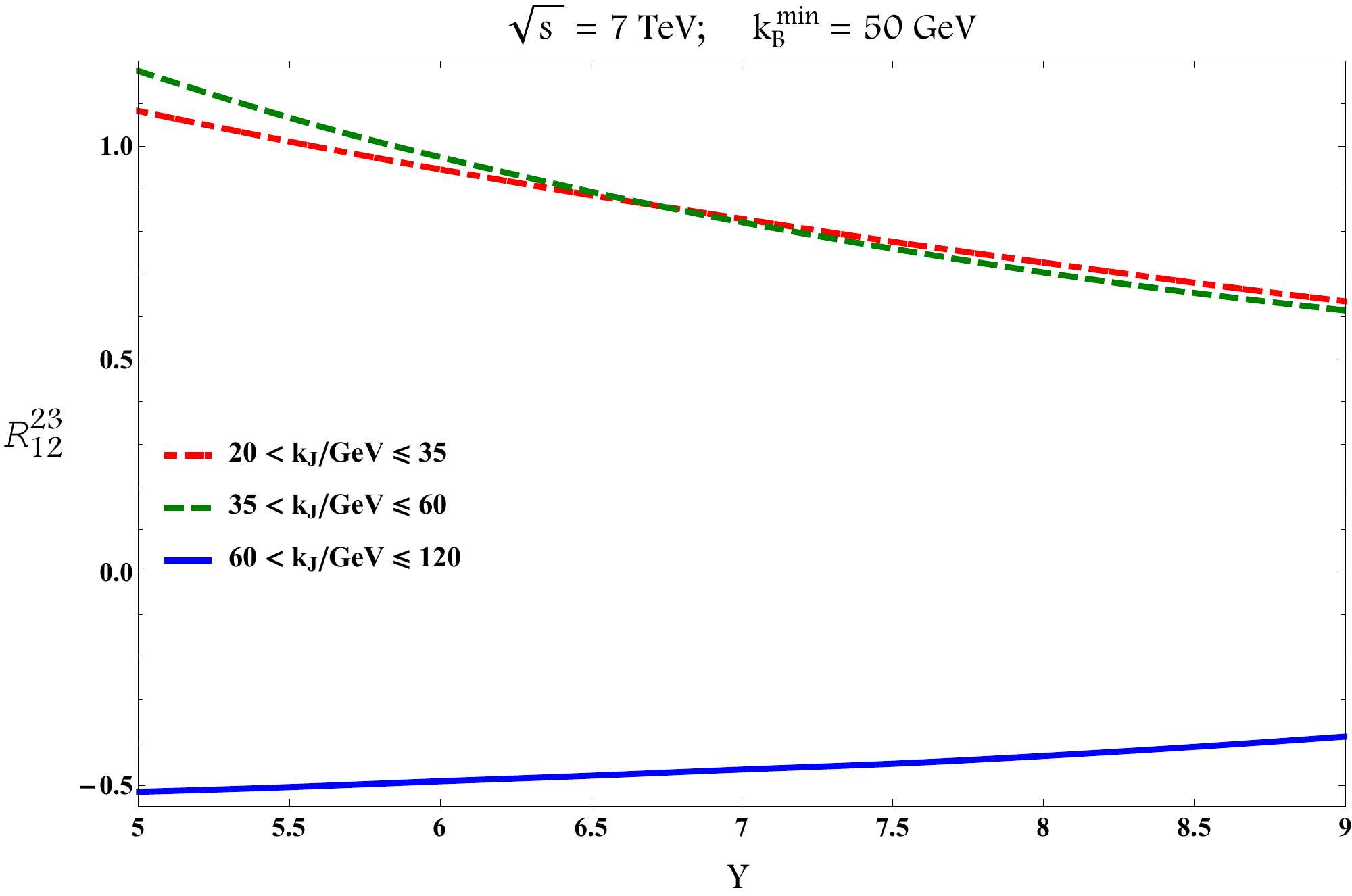}
   \vspace{1cm}

   \hspace{-2.25cm}
   \includegraphics[scale=0.45]{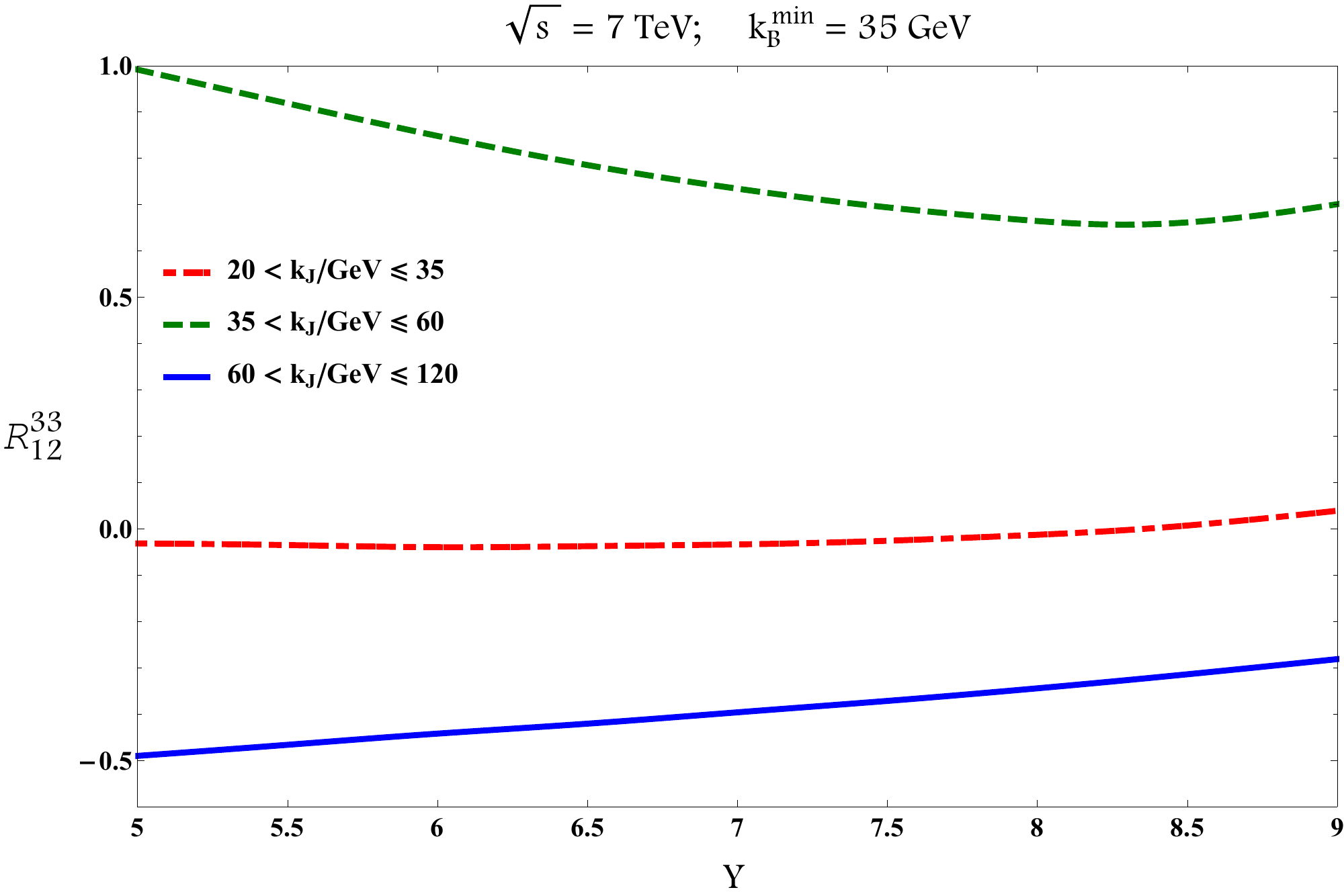}
   \includegraphics[scale=0.45]{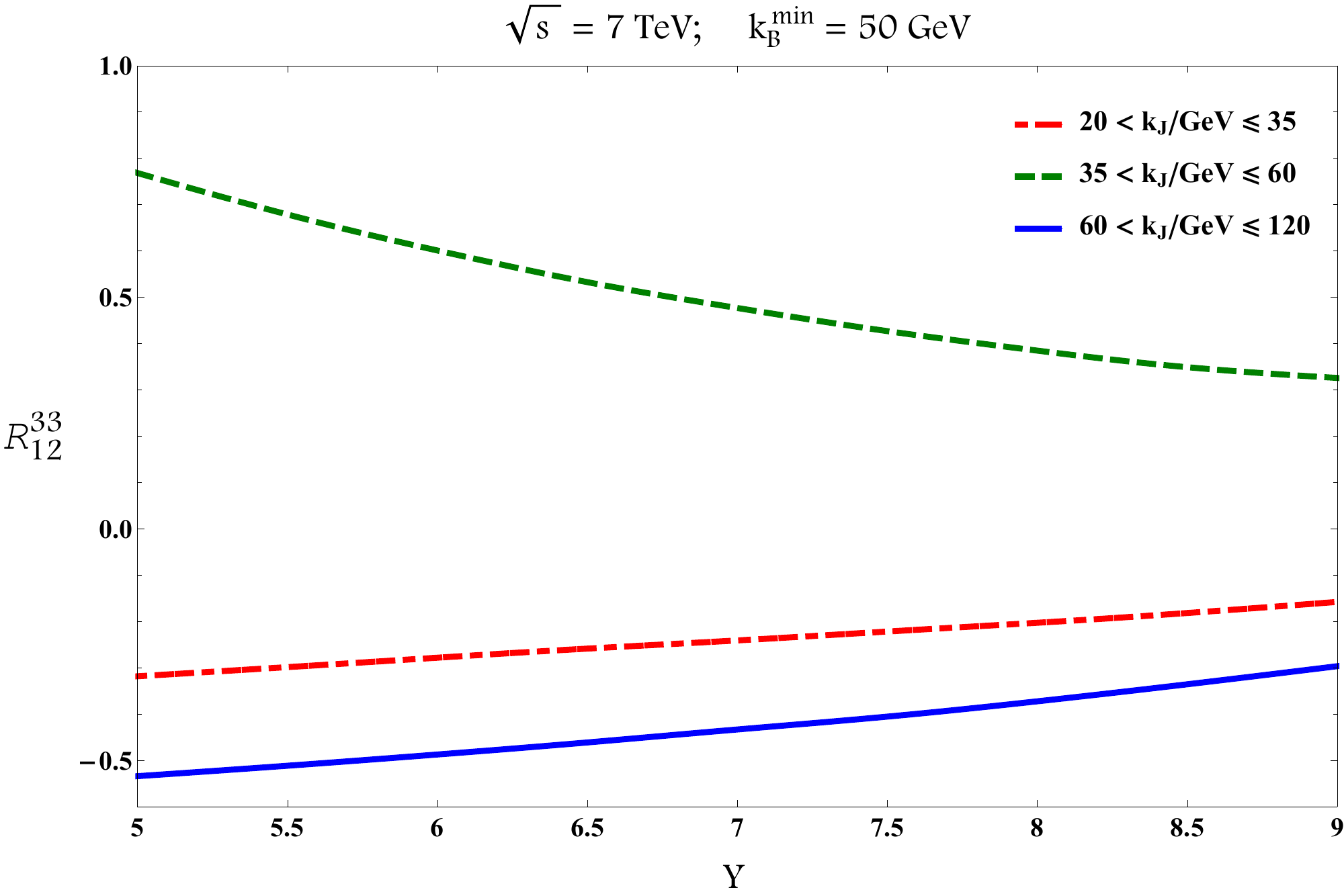}
   \vspace{1cm}

   \hspace{-2.25cm}   
   \includegraphics[scale=0.45]{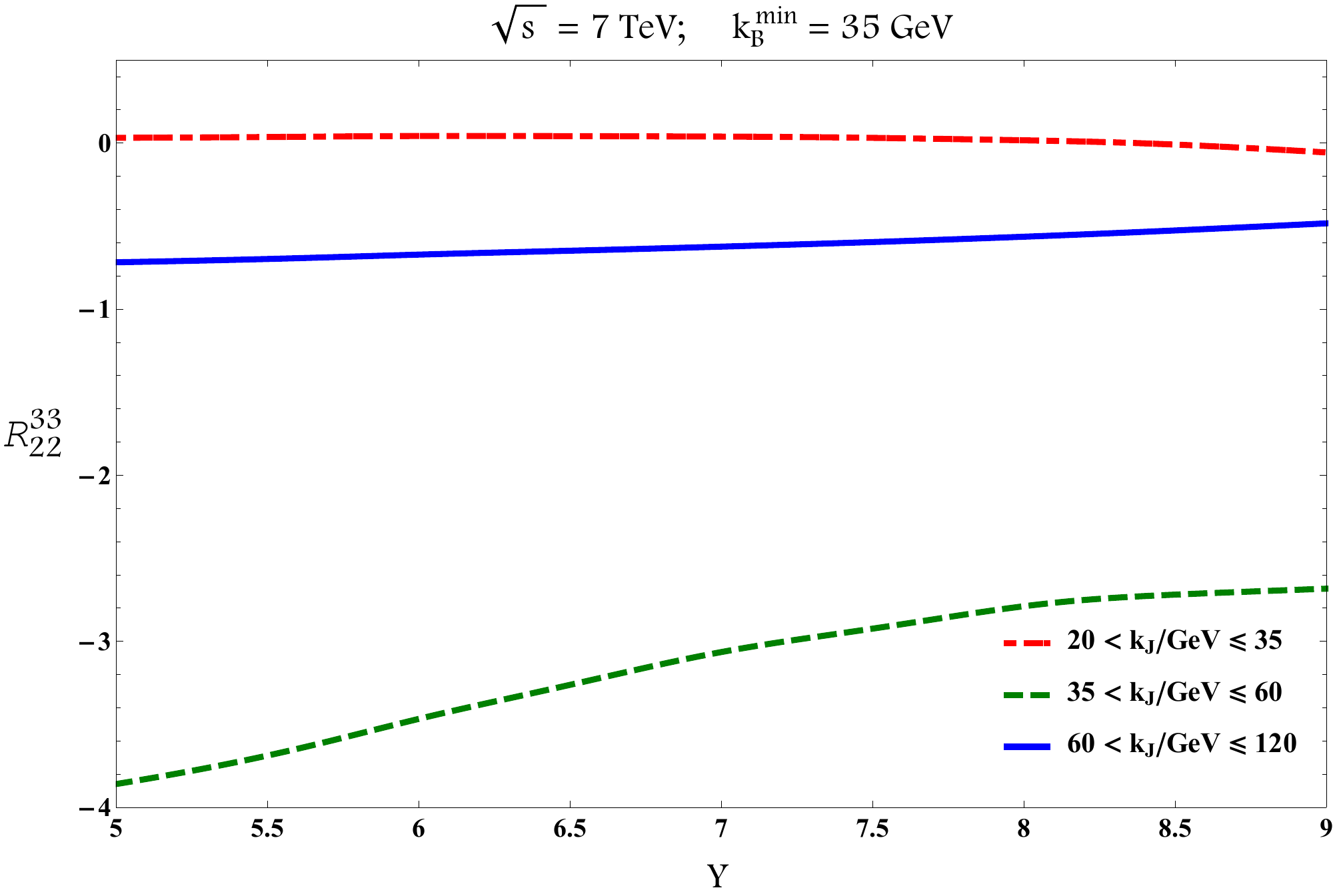}
   \includegraphics[scale=0.45]{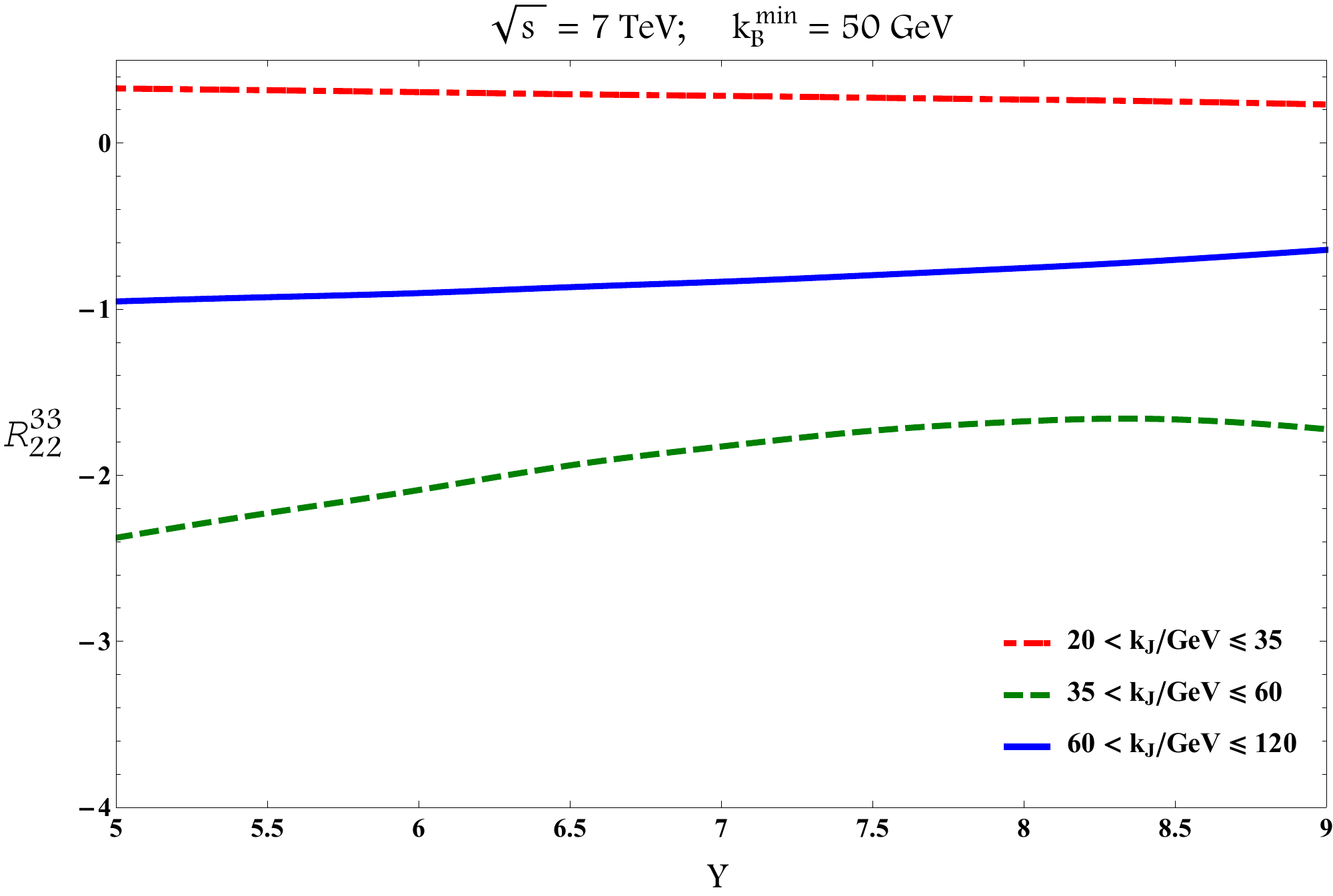}

\restoregeometry
\caption{\small $Y$-dependence of 
$R^{23}_{12}$, $R^{33}_{12}$ and $R^{33}_{22}$ 
for $\sqrt s = 7$ TeV and $k_B^{\rm min} = 35$ GeV (left column)
and $k_B^{\rm min} = 50$ GeV (right column).} 
\label{fig:7-second}
\end{figure}

\begin{figure}[p]
\newgeometry{left=-10cm,right=1cm}
\vspace{-2cm}

   \hspace{-2.25cm}
   \includegraphics[scale=0.45]{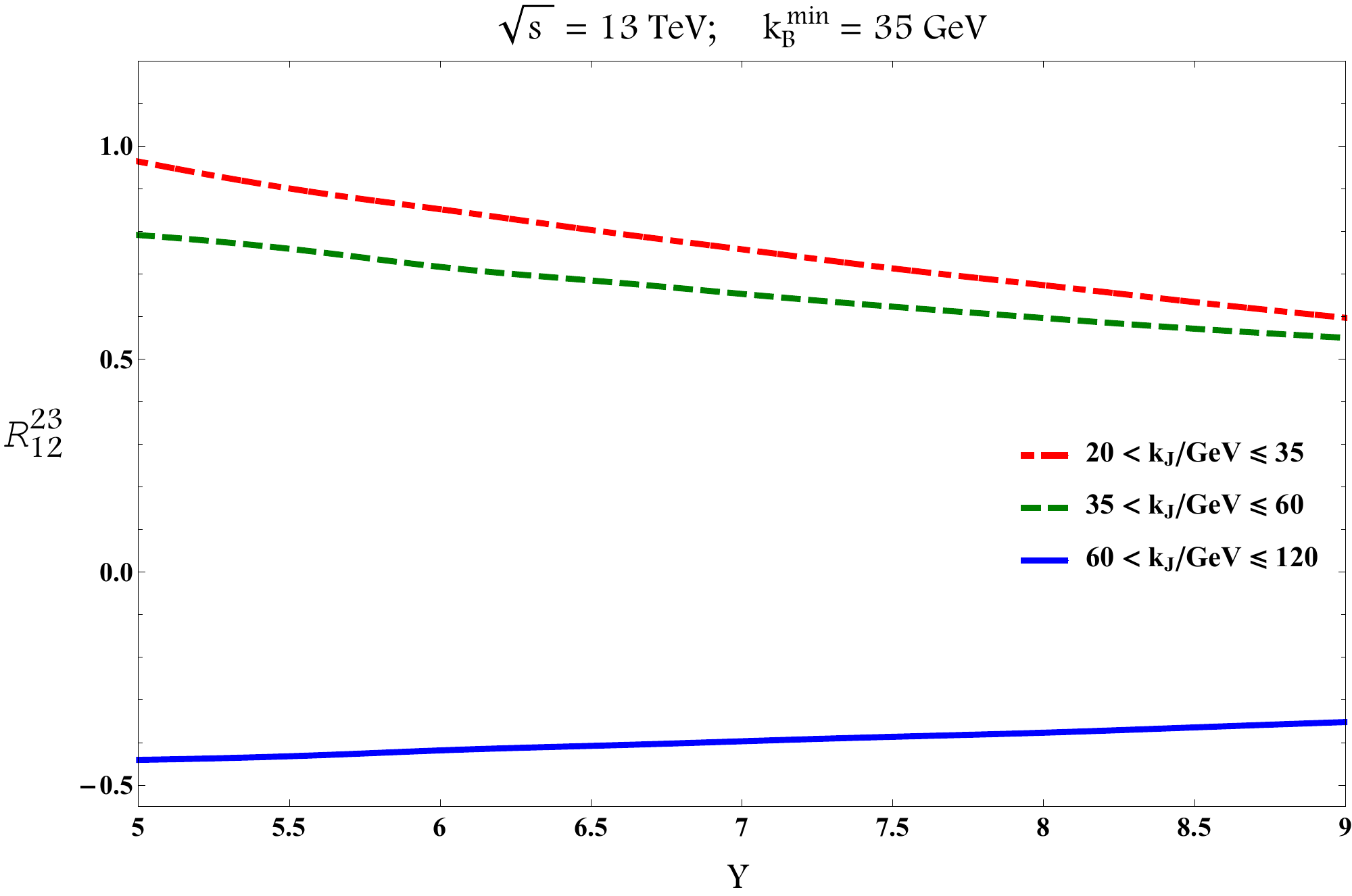}
   \includegraphics[scale=0.45]{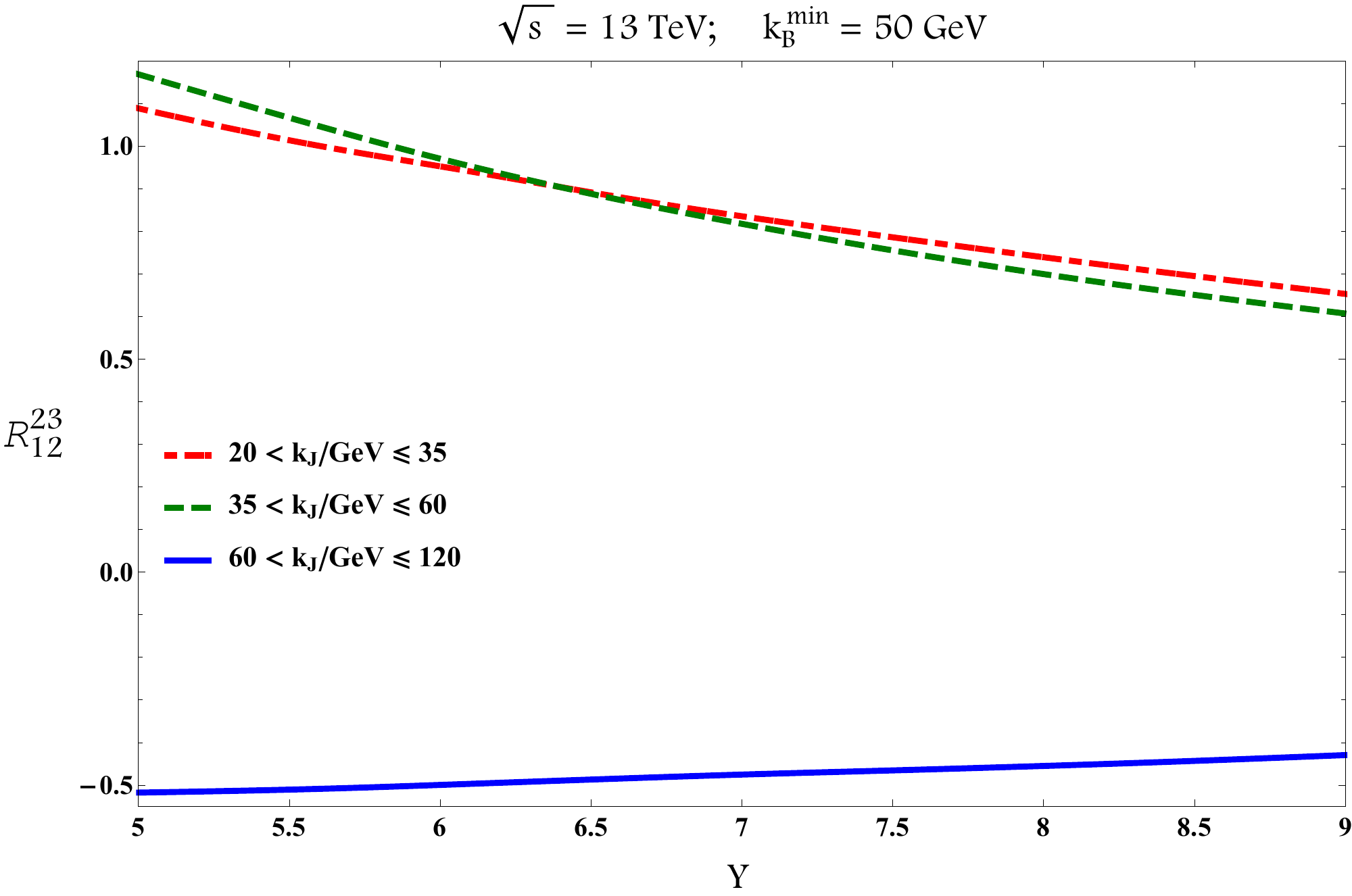}
   \vspace{1cm}

   \hspace{-2.25cm}
   \includegraphics[scale=0.45]{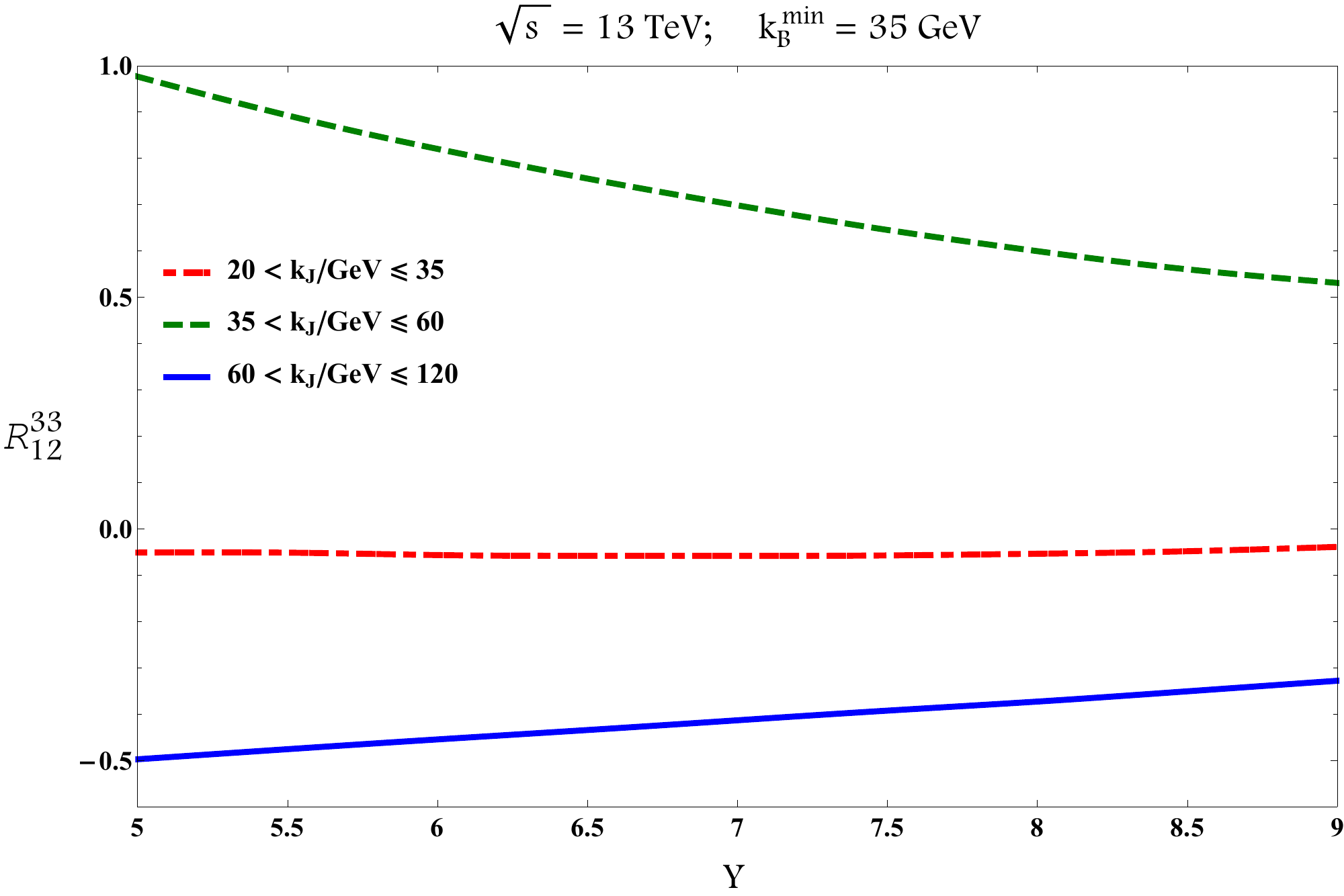}
   \includegraphics[scale=0.45]{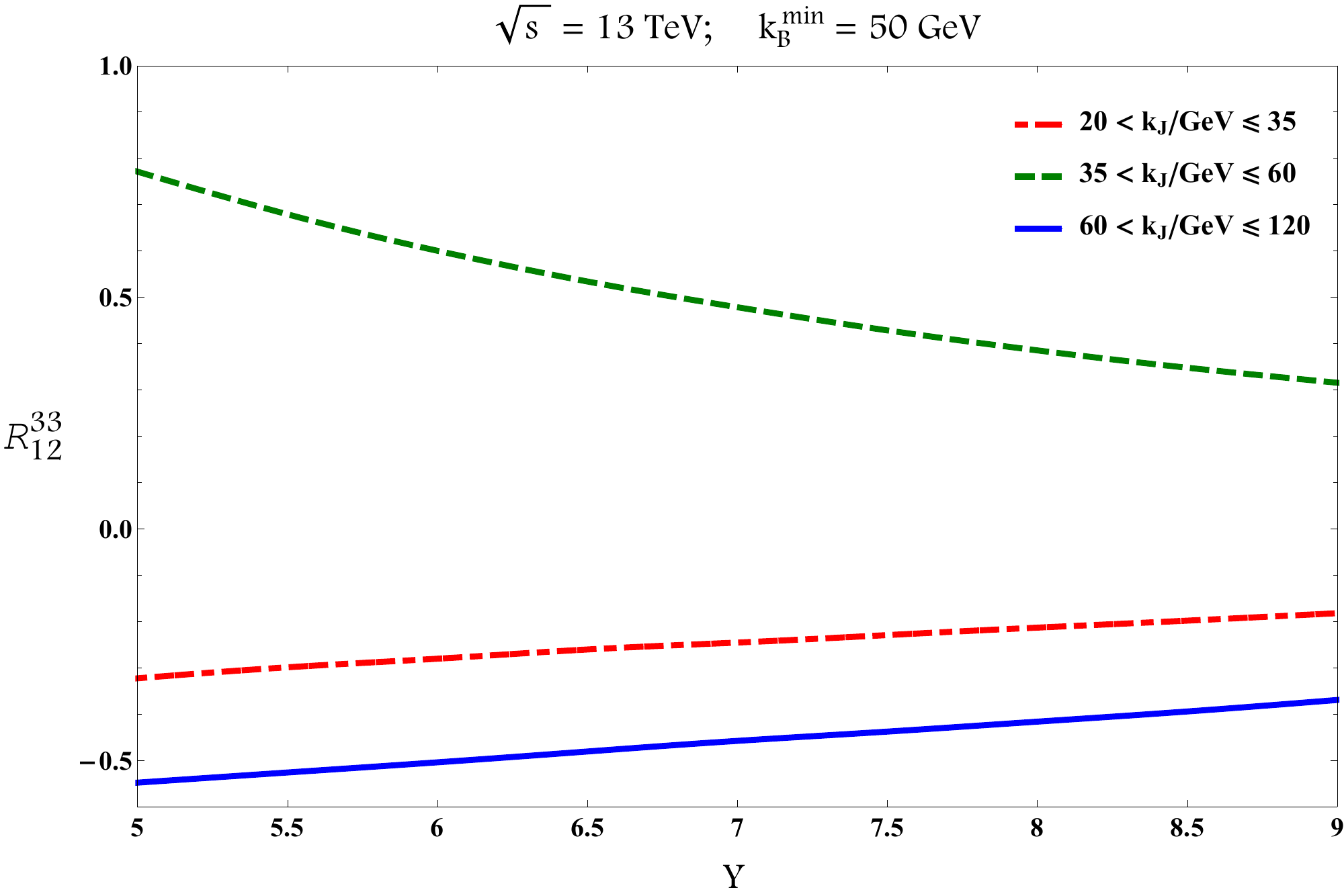}
   \vspace{1cm}

   \hspace{-2.25cm}   
   \includegraphics[scale=0.45]{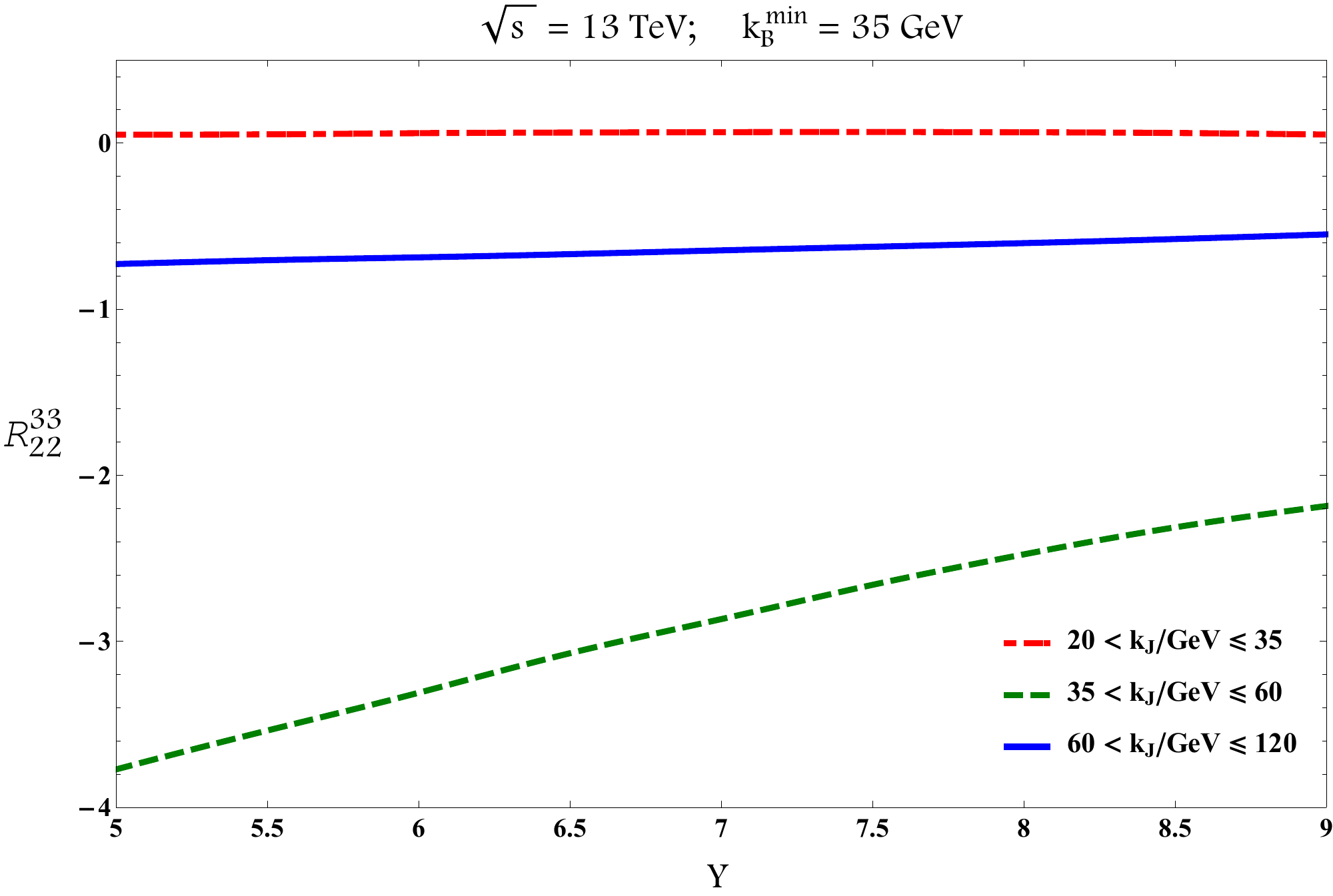}
   \includegraphics[scale=0.45]{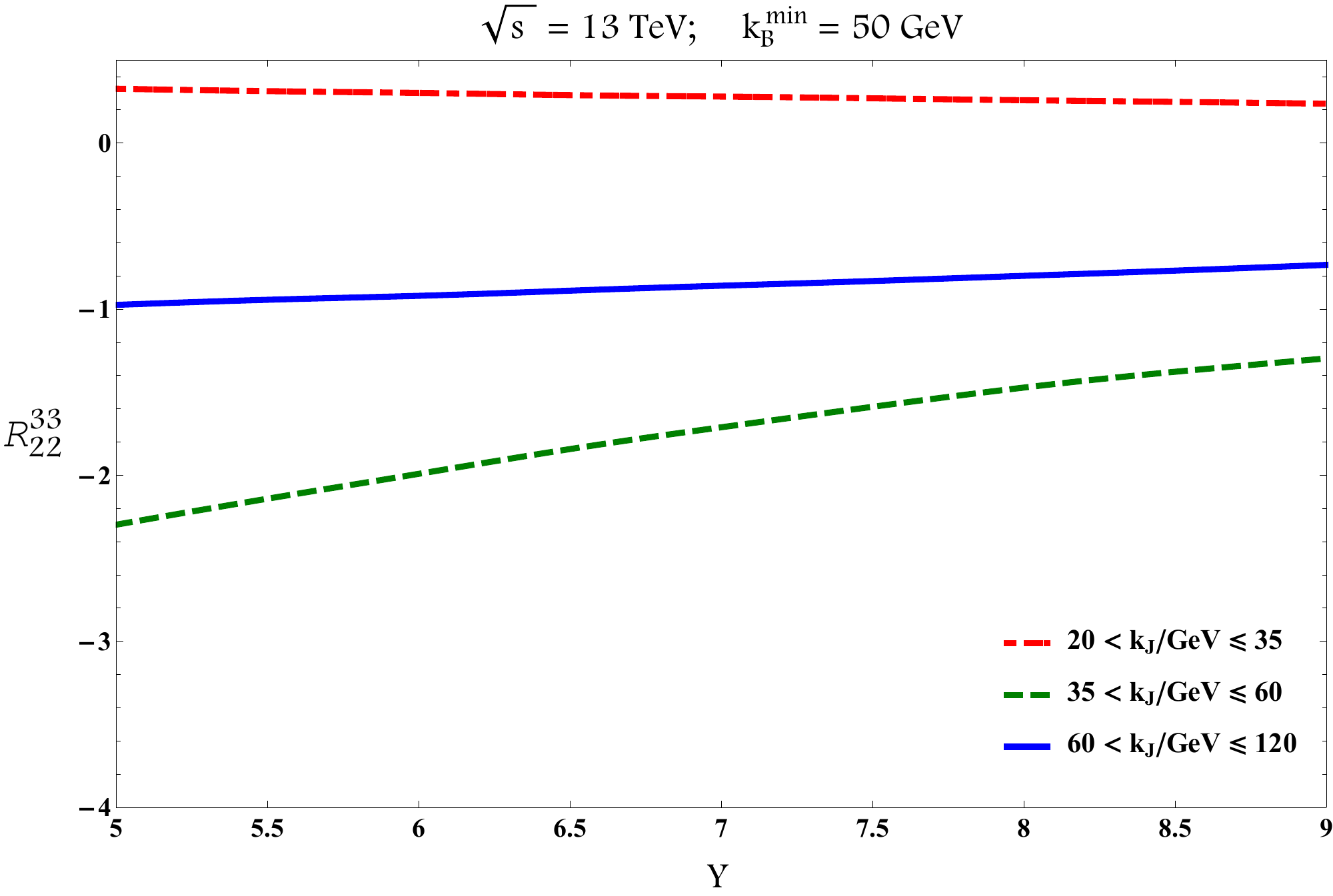}

\restoregeometry
\caption{\small $Y$-dependence of 
$R^{23}_{12}$, $R^{33}_{12}$ and $R^{33}_{22}$ 
for $\sqrt s = 13$ TeV and $k_B^{\rm min} = 35$ GeV (left column)
and $k_B^{\rm min} = 50$ GeV (right column).} 
\label{fig:13-second}
\end{figure}

In the following, we present our results collectively in four figures.
In Figs.~\ref{fig:7-first} and \ref{fig:7-second} different ratios
are shown for $\sqrt{s} = 7$ TeV and in Figs.~\ref{fig:13-first} and \ref{fig:13-second}
we see the same ratios for  $\sqrt{s} = 13$ TeV.
In all four figures, in the left column we place the plots for the symmetric kinematic cut
($k_B^{\rm min} = 35$ GeV) and in the right column the plots for the asymmetric one
($k_B^{\rm min} = 50$ GeV). The red dot-dashed
curve corresponds to $k_J$  bounded in bin-1,
the green dashed curve to $k_J$  bounded in bin-2 and finally the blue 
continuous one to $k_J$ bounded in bin-3. In total, we show the results
for six different observables: 
$R_{22}^{11}$, $R_{12}^{13}$, $R_{12}^{22}$,
$R_{12}^{23}$, $R_{12}^{33}$ and $R_{22}^{33}$.

The first observation that becomes apparent from a preliminary view to the four
figures is that the dependence of the different observables on the rapidity
difference between $k_A$ and $k_B$ is rather smooth. This is more
pronounced when we consider $k_J$ being larger than the
forward/backward jets (blue line). Indeed, the blue curve, which corresponds to large values of the transverse momentum in the central jet, 
is mostly linear. The other two curves (red and green) follow
generally the same smooth with Y behavior although 
they tend to be less linear than the blue curve.

The slope of the three curves, in absolute values, depends
on the particular observable. For example, in Fig.~\ref{fig:7-first}, 
the blue curve in the top left drops from $\sim 3.5$ at $Y=5$
to $\sim 4.5$ at $Y=9$, whereas in bottom left it drops from
$\sim 0.7$ to $\sim 0.6$. 

Another interesting observation is that there are ratios for which
changing from the symmetric to the asymmetric cut makes no real difference
and other ratios for which the picture changes radically.
A characteristic example of the former case 
is the observable $R_{12}^{22}$ in Fig.~\ref{fig:7-first} bottom line, where we 
see practically no big differences between the left and right plots.
If instead we focus on $R_{12}^{13}$ in Fig.~\ref{fig:7-first} middle line, we see
that going from the symmetric cut (left) to the asymmetric one
(right) brings forward a big change.

The main conclusion we would like to draw 
after comparing Fig.~\ref{fig:7-first} to Fig.~\ref{fig:13-first}
and  Fig.~\ref{fig:7-second} to Fig.~\ref{fig:13-second} is that, in general, 
for most of the observables there are no
significant changes when we increase the colliding energy from
7 to 13 TeV. This is indeed remarkable since it indicates that a sort of asymptotic regime has been reached for the kinematical configurations included in our analysis. It also tells us that our observables
are really as insensitive as possible to effects
which have their origin outside the BFKL dynamics and which 
normally cannot be isolated  ({\it e.g.} influence from the PDFs).

\section{Summary \& Outlook}

We have presented a first full phenomenological study of inclusive
three-jet production at the LHC within the BFKL framework, focussing on the study of azimuthal-angle dependent observables.
Following the work done in Ref.~\cite{Caporale:2015vya} where
a new family of observables was proposed  
to probe the window of applicability of BFKL at the LHC, 
we have studied here
a selection of these observables at a hadronic level (with PDFs)
at two different colliding
energies, $\sqrt s = 7, 13$ TeV. We have considered a symmetric 
and an asymmetric kinematic cut with respect to the transverse momentum
of the forward ($k_A$) and backward ($k_B$) jets. 
In addition, we have chosen to impose an
extra condition on the value of the transverse momentum $k_J$ of the central jet,
dividing the allowed region for $k_J$ into three sub-regions: $k_J$ smaller than
$k_{A,B}$, $k_J$ similar to $k_{A,B}$ and $k_J$ larger than $k_{A,B}$.

We have shown how our observables 
$R_{22}^{11}$, $R_{12}^{13}$, $R_{12}^{22}$,
$R_{12}^{23}$, $R_{12}^{33}$ and $R_{22}^{33}$
change when we vary the rapidity
difference Y between $k_A$ and $k_B$ from 5 to 9 units.
We notice a generally smooth functional dependence of
the ratios on Y. These observables do not considerably change when we increase
the colliding energy from 7 to 13 TeV which assures us that
they capture the essence of what the BFKL dynamics dictates regarding
the azimuthal behavior of the hard jets in inclusive three-jet production.
It will be very interesting to compare with possible predictions for these observables from 
fixed order analyses as well as from the BFKL inspired 
Monte Carlo \cod {BFKLex}~\cite{Chachamis:2011rw,Chachamis:2011nz,Chachamis:2012fk,
Chachamis:2012qw,Caporale:2013bva,Chachamis:2015zzp,Chachamis:2015ico}. Predictions from general-purpose Monte Carlos should also be put forward. 

Most importantly though, it would be extremely interesting to see an
experimental analysis for these observables using the existing and future LHC data.
We would like to motivate our experimental colleagues to proceed to such
an analysis since we believe it will help address  
the question of how phenomenologically
relevant the BFKL dynamics is at present energies. It would also serve as a very good test of models describing multiple interactions and to gauge how important those effects can be.

\begin{flushleft}
{\bf \large Acknowledgements}
\end{flushleft}
GC acknowledges support from the MICINN, Spain, 
under contract FPA2013-44773-P. 
ASV acknowledges support from Spanish Government 
(MICINN (FPA2010-17747,FPA2012-32828)) and, together with FC and FGC, 
to the Spanish MINECO Centro de Excelencia Severo Ochoa Programme 
(SEV-2012-0249). FGC thanks the Instituto de F{\'\i}sica Te{\'o}rica 
(IFT UAM-CSIC) in Madrid for warm hospitality.

\end{document}